\documentclass[
aps,pra,reprint,superscriptaddress,
nofootinbib,nobibnotes,floatfix
]{revtex4-2}

\usepackage[T1]{fontenc}
\usepackage[utf8]{inputenc}
\usepackage[british]{babel}
\usepackage{mathpazo}
\usepackage{comment}
\usepackage{tikz}
\usepackage{microtype}
\usepackage{amsmath,amssymb,amsthm,mathtools,booktabs,graphicx}
\usepackage[dvipsnames]{xcolor}
\usepackage[colorlinks=true,citecolor=Magenta,urlcolor=Blue,linkcolor=Blue]{hyperref}
\graphicspath{{figures/}}
\newcommand{\Fp}{\mathbb F_p}
\newcommand{\Mn}{\mathcal M_n}
\DeclareMathOperator{\Tr}{Tr}
\DeclareMathOperator{\supp}{supp}
\newcommand{\E}{\mathbb E}
\newcommand{\NC}{\mathrm{NC}}
\newcommand{\PS}{\mathrm{PS}}
\newcommand{\excl}{\mathrm{excl}}
\newcommand{\ret}{\mathrm{ret}}
\newcommand{\ket}[1]{\lvert #1\rangle}
\newcommand{\bra}[1]{\langle #1\rvert}
\newcommand{\ind}{\mathbf 1}

\newtheorem{theorem}{Theorem}
\newtheorem{corollary}{Corollary}
\newtheorem{proposition}{Proposition}
\newtheorem{lemma}{Lemma}

\begin{document}
\title{Exponential Quantum Separation of Retrieval and Exclusion with All-versus-Nothing Preparation Contextuality}
\author{Pritam Roy}
\email{roy.pritamphy@gmail.com}
\affiliation{S. N. Bose National Centre for Basic Sciences, Block JD, Sector III, Salt Lake, Kolkata 700106, India}

\begin{abstract}
Knowing a matched difference can be exponentially more expensive
than conclusively excluding one of its possible values.
We establish this fundamental asymmetry in quantum communication
through Difference Hidden Matching and Antimatching over
odd-prime symbols.
Although both tasks require $\Theta_p(n)$ classical bits for
exact communication, retrieving the correct difference requires
$\Theta_p(n)$ qubits for arbitrary encodings, whereas perfect
exclusion needs only $\log_2 n+O_p(1)$ qubits.
An explicit phase encoding achieves exclusion through destructive
interference, while an encoding-independent bound establishes
its optimal leading logarithmic cost.
Remarkably, the same interference-based construction also reveals
an all-versus-nothing contradiction with preparation noncontextuality
under pair-difference parity obliviousness.
We prove that positivity confines noncontextual conclusive
exclusions to the edges of a single triangle, yielding a complete
characterization of matching families admitting perfect noncontextual exclusion and a minimal
three-setting witness with tight odd-prime noncontextual bound
$1-1/p^2$, against unit quantum success.
For characterized ternary preparations, the exclusion statistics
further identify the antitrine POVM and witness measurements
beyond projective simulability.
Thus, quantum exclusion provides both an exponential saving in
communication resources and an operational probe of nonclassicality.
\end{abstract}
\maketitle

\textit{Introduction.}---
What can a quantum message reveal about an unknown preparation?
State discrimination identifies a possibility~\cite{Helstrom1976},
whereas conclusive exclusion certifies which possibility did
\emph{not} occur. For binary alternatives these are equivalent,
but beyond two symbols nonorthogonal states may be
antidistinguished even when they cannot be perfectly
discriminated~\cite{bandyopadhyay,heinosaari,PBR2012,Johnston2025}.
This distinction has consequences for preparation
contextuality~\cite{leifer,srikumar,ying}, resource theories
of exclusion~\cite{Uola2020}, communication
advantages~\cite{havlicek,bae}, and generalized
measurements~\cite{Diebra2026,YaoWang2026}.
Exclusion games have already produced unbounded quantum--classical
information-cost gaps, while the exhibited zero-error protocol sends $n$ qubits~\cite{Perry2015Exclusion}.
This leaves open a different question: whether \emph{retrieving and excluding
an answer to the same distributed problem} can require
exponentially different optimal quantum message dimensions.

One-way relational communication is a natural setting for this
question because Alice must encode information before learning
which relation Bob will query. The Holevo bound limits
accessible classical information~\cite{Holevo73},
while quantum random access codes quantify how
well compressed messages recover designated
symbols~\cite{Nayak1999,Ambainis2002,
Tavakoli2015QRAC,Ambainis2024QRAC}.
Relational communication instead permits Bob
to choose among several valid outputs~\cite{Emeriau2022}.
 In binary Hidden
Matching (HM), Alice holds an $n$-bit string and sends a
message before learning Bob's perfect matching; Bob reports
the parity of \emph{any} matched pair. This problem admits
$O(\log n)$-qubit communication, while a fixed classical
advantage requires $\Omega(\sqrt n)$ bits~\cite{byjk}.
Matching problems have become useful beyond that separation:
they inform further communication-complexity
bounds~\cite{doriguello,bun}, bounded-storage cryptographic
separations~\cite{gavinsky,chakraborty}, graph-streaming
memory lower bounds~\cite{Kapralov2022NoisyBHM}, and optical protocol constructions~\cite{arrazola}
and an experimental demonstration of Sampling
Matching~\cite{kumar}.
These applications motivate a more basic resource question:
what relational information can a short quantum message
certify when Bob's query is unknown to Alice?
Binary HM cannot distinguish retrieval from exclusion,
because excluding one parity determines the other.

Beyond communication cost, the distinction between retrieval and
exclusion raises a foundational question: under what operational
restrictions can perfect relational exclusion be explained by a
preparation-noncontextual model? Such a model assigns the same hidden-state distribution
to preparation mixtures that no measurement can
distinguish~\cite{spekkens,SchmidEquv2018}, a notion distinct
from traditional Kochen--Specker contextuality~\cite{Bell1966,Kochen1967}.
Parity-oblivious communication provides a concrete way
to enforce these operational equivalences and test their
consequences~\cite{porac,hameedi,ghorai,Chailloux2016PORAC,Ambainis2019dPORAC,Roy2024GPOC}.
Related communication advantages have been connected
to preparation contextuality~\cite{Sahacommunicationadvantage2019,Gupta2023},
including conclusive exclusion~\cite{ying} and parity-oblivious random
exclusion codes~\cite{porec}.
Recently, binary parity-oblivious Hidden Matching (PoHM)
yielded an all-versus-nothing (AVN) contradiction and
a triangle restriction on perfectly certifiable pair
parities~\cite{pohm}.
Does this foundational obstruction survive when retrieval
and exclusion are genuinely different tasks?

\begin{figure*}[t]
\centering
\includegraphics[
    width=0.7\textwidth,
    keepaspectratio
]{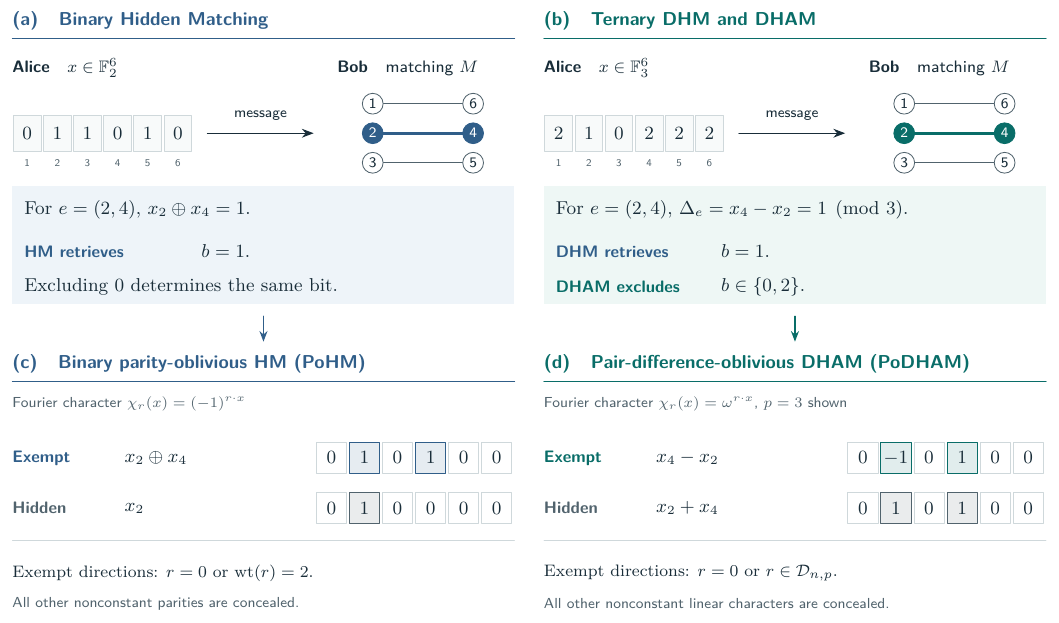}
\caption{\footnotesize
Hidden Matching and its parity-oblivious extensions.
Alice sends a one-way message without knowing Bob's
perfect matching $M$.
(a) Binary HM: retrieving a matched parity is
equivalent to excluding its opposite.
(b) Ternary DHM requires the true matched difference,
whereas DHAM accepts any incorrect value; for
$\Delta_{24}=1$, the respective outputs are
$b=1$ and $b\in\{0,2\}$.
(c) Binary PoHM permits pair-parity Fourier modes
while concealing other nonconstant modes.
(d) Odd-prime PoDHAM similarly permits pair-difference
modes $a(\mathbf e_j-\mathbf e_i)$ while concealing
all other nonconstant modes. Permitted modes
need not be present in a particular encoding.
}
\label{fig:tasks}
\end{figure*}
\textit{In this Work}, we establish an exponential
separation between the quantum resources needed to
retrieve and exclude the same relational information.
We introduce Difference Hidden Matching (DHM) and
Difference Hidden Antimatching (DHAM) over odd-prime
symbols ($p$). Both require $\Theta_p(n)$ bits of exact
classical communication, yet exact quantum retrieval
requires $\Theta_p(n)$ qubits while exclusion costs only
$\log_2 n+O_p(1)$ qubits. Encoding-independent bounds
establish the optimal asymptotic costs, and an explicit
phase protocol achieves perfect exclusion by destructive
interference. For this encoding and uniformly distributed inputs,
matched-edge retrieval succeeds with probability at most $2/p$, illustrating
how exclusion can succeed without identification.

Under pair-difference parity obliviousness, the same
protocol gives an all-versus-nothing demonstration of
preparation contextuality. Positivity of noncontextual
ontic posteriors imposes a triangle obstruction that
characterizes matching families admitting perfect
noncontextual exclusion and yields the tight
three-cyclic-setting bound $1-1/p^2$ for odd primes,
against unit quantum success. We also determine the
full-family asymptotic noncontextual scaling and show
that ternary exclusion statistics certify
nonprojective-simulable qubit measurements under
characterized-source, edge-resolved assumptions.

\textit{Difference Hidden Matching and Antimatching.}---
To distinguish retrieval from exclusion within the same relational
communication problem, we extend binary Hidden Matching to an
odd-prime alphabet $\Fp$. The matching determines which differences
Bob may report, but the two tasks impose different conditions on his
output.

Let $n\ge4$ be even. Alice receives
$x=(x_1,\ldots,x_n)\in\Fp^n$, while Bob receives a
perfect matching $M\in\Mn$ of $[n]$, unknown to Alice.
A perfect matching pairs all $n$ positions, so $n$ must
be even, and $n=4$ is the smallest size admitting more
than one pairing.
Orienting each edge $e=(i,j)$ with $i<j$, we define
\begin{equation}
\Delta_e(x)=x_j-x_i\pmod p.
\label{eq:diff}
\end{equation}
After receiving a single message from Alice, Bob
selects any edge $e\in M$ and outputs $b\in\Fp$.
In Difference Hidden Matching ($\mathrm{DHM}_{n,p}$),
he must retrieve $b=\Delta_e(x)$; in Difference Hidden
Antimatching ($\mathrm{DHAM}_{n,p}$), he must report a
value $b\ne\Delta_e(x)$, thereby ruling out $b$.
Fig.~\ref{fig:tasks}(a) and (b) illustrate binary Hidden
Matching and the nonbinary retrieval--exclusion distinction.

For binary symbols, $x_i+x_j=x_j-x_i\pmod 2$, whereas sums
and differences define distinct tasks for odd $p$. We choose
differences because they are invariant under a common translation
$x\mapsto x+c\mathbf1$ and satisfy
$\Delta_{ij}+\Delta_{jk}=\Delta_{ik}$ for $i<j<k$.
The first property is suited to encoding information in relative
phases, while the second captures the consistency of differences
around a triangle. The $p$-ary Hidden Matching decision problem
studied in streaming lower bounds has a different objective and
input promise~\cite{guruswami}.

We consider one-way protocols with public randomness, arbitrary
mixed-state encodings and POVMs, and no preshared entanglement.
When public randomness is used, Bob has access to both the
revealed seed and the transmitted message.
A protocol achieves zero error if Bob returns a valid pair
$(e,b)$ with certainty for every input $x$ and matching $M$.

\textit{Classical communication benchmark.}---
We first establish the classical cost of the two tasks. Although
exclusion permits $p-1$ possible labels on each edge, this freedom
does not reduce the asymptotic communication cost.

\begin{theorem}[Classical communication]\label{thm:classical}
For every fixed prime $p$,
\begin{equation}
C_0^\rightarrow(\mathrm{DHM}_{n,p})=\Theta_p(n),\qquad
C_0^\rightarrow(\mathrm{DHAM}_{n,p})=\Theta_p(n).
\label{eq:classical}
\end{equation}
For any fixed positive advantage above the respective random-guessing probabilities and strictly below perfect success, both public-coin costs are $\Theta_{p,\eta}(\sqrt n)$.
\end{theorem}
To see why even exclusion requires linear communication, associate
with each deterministic message a graph containing those edges on
which at least one difference can be excluded for every input
consistent with that message. Exactness requires this graph to meet
every perfect matching, which forces a connected component of size
greater than $n/2$. A spanning-tree count then bounds the number of
inputs compatible with a single message. The retrieval lower bound
follows because an exact retrieval protocol also solves exclusion.
Conversely, communicating the relative values of $n/2+1$ positions
suffices for both tasks. In the bounded-error regime, public sampling
and finite-field Fourier bounds yield the matching $\sqrt n$ scaling.
The proofs are given in Appendix~\ref{app:classical}.

\textit{Exact quantum separation.}---
The classical equivalence in asymptotic cost breaks down when the
message is quantum. Relative phases allow Bob to rule out a matched
difference without determining it, leading to the following separation.
\begin{theorem}[Exact quantum separation]\label{thm:quantum}
For every fixed odd prime $p$ and even $n\ge4$,
the exact unassisted one-way quantum communication costs satisfy
\begin{equation}
\begin{aligned}
Q_0^\rightarrow(\mathrm{DHM}_{n,p})&=\Theta_p(n),\\
Q_0^\rightarrow(\mathrm{DHAM}_{n,p})&=\log_2 n+O_p(1).
\end{aligned}
\label{eq:quantum-cost}
\end{equation}
\end{theorem}
More explicitly, the retrieval cost and the minimum message dimension
$d_{\min}$ for exact exclusion satisfy
\begin{align}
Q_0^\rightarrow(\mathrm{DHM}_{n,p})
&\ge n\log_2\frac{p}{2\sqrt{p-1}}-\log_2(2p),
\label{eq:qlower}\\
1+\frac{n-1}{p-1}&\le d_{\min}(\mathrm{DHAM}_{n,p})\le n.
\label{eq:qudimension}
\end{align}
We first establish the exclusion upper bound by encoding Alice's
symbols in the relative phases of an $n$-dimensional state,
\begin{equation}
\ket{\psi_x}=\frac1{\sqrt n}\sum_{j=1}^n\omega^{x_j}\ket j,
\qquad \omega=e^{2\pi i/p}.
\label{eq:phase}
\end{equation}
This encoding realizes the translation invariance of the task:
its density operator $\rho_x=\ket{\psi_x}\bra{\psi_x}$ satisfies
\begin{equation}
\langle i|\rho_x|j\rangle=\omega^{x_i-x_j}/n,
\qquad\rho_{x+c\mathbf1}=\rho_x.
\label{eq:em-coherence}
\end{equation}
In contrast, a pair sum changes as $x_i+x_j\mapsto x_i+x_j+2c$.
Since $2$ is invertible for odd $p$, every sum label occurs among
inputs represented by the same state. Thus this particular ensemble
cannot achieve perfect sum exclusion. Other encodings are not ruled
out; for a bipartite matching family, sign changes on one vertex
class convert sums into differences.

To exploit the relative phase on a matched edge $e=(i,j)$, Bob
uses the effects
\begin{equation}
E_{e,b}=\frac2p\ket{\chi_{e,b}}\bra{\chi_{e,b}},\quad
\ket{\chi_{e,b}}=\frac{\ket i-\omega^b\ket j}{\sqrt2}.
\label{eq:povm}
\end{equation}
For each edge, $\sum_b E_{e,b}=\ket i\bra i+\ket j\bra j$.
Since the matching partitions the basis, the complete collection
of effects forms a POVM. Its outcome probability is
\begin{equation}
\Pr(e,b\mid x,M)=\frac1{np}
\bigl|1-\omega^{\Delta_e(x)-b}\bigr|^2,
\label{eq:prob}
\end{equation}
which vanishes whenever $b=\Delta_e(x)$. Every outcome therefore
identifies an incorrect difference, using only
$\lceil\log_2 n\rceil$ qubits.

Why does the analogous compression fail for exact retrieval? If two
inputs admit no common valid output for some matching, their quantum
messages must have orthogonal supports. Appendix~\ref{app:quantum}
constructs exponentially many pairwise incompatible inputs and
thereby establishes Eq.~\eqref{eq:qlower}. Sending the entire string
provides the matching linear upper bound.

It remains to determine whether exclusion can be achieved with a
substantially smaller message dimension than the phase protocol.
The key observation is that the dimension limits the total Fourier
weight of an arbitrary encoding, whereas exact exclusion requires
nonzero weight on the edges of every matching. For an arbitrary
exact $d$-dimensional DHAM encoding $x\mapsto\rho_x$, define
\begin{equation}
\begin{gathered}
\bar\rho=\E_x\rho_x,\quad R_r=\E_x[\omega^{-r\cdot x}\rho_x],\\
w_r=\Tr(R_r^\dagger\bar\rho^{-1}R_r),
\end{gathered}
\label{eq:em-weight}
\end{equation}
where $r\in\Fp^n$, the expectation is uniform over inputs, and
the inverse acts on $\supp\bar\rho$, of rank $s\le d$.
Matrix-valued Parseval and $\rho_x^2\le\rho_x$ imply
\begin{equation}
\sum_r w_r=\E_x\Tr(\rho_x\bar\rho^{-1}\rho_x)\le s.
\label{eq:em-parseval}
\end{equation}
Since $R_0=\bar\rho$, we have $w_0=1$ and hence
$\sum_{r\ne0}w_r\le s-1$.
This bounds the total available nonconstant Fourier weight.
To bound the amount required for exclusion, consider a fixed
matching $M$. Exactness forces the sum of the probabilities
of all outcomes reporting the true difference to vanish. Expanding
their indicators in additive characters, the constant Fourier
contribution must cancel the nonconstant terms. Applying
Hilbert--Schmidt and scalar Cauchy--Schwarz inequalities,
together with the matching-dependent POVM normalization, gives
\begin{equation}
1\le(p-1)\sum_{e=(i,j)\in M}\sum_{a\in\Fp^*}
w_{a(\mathbf e_j-\mathbf e_i)}.
\label{eq:em-matching}
\end{equation}
Every edge occurs in a uniformly random perfect matching with
probability $1/(n-1)$, and different edges determine disjoint
nonzero Fourier lines. Averaging Eq.~\eqref{eq:em-matching}
therefore yields
\begin{equation}
\begin{aligned}
1&\le\frac{p-1}{n-1}\sum_{i<j}\sum_{a\ne0}
 w_{a(\mathbf e_j-\mathbf e_i)}\\
&\le\frac{(p-1)(s-1)}{n-1}.
\end{aligned}
\label{eq:em-average}
\end{equation}
Consequently, $d\ge s\ge1+(n-1)/(p-1)$, establishing the lower
bound in Eq.~\eqref{eq:qudimension}. No restriction on the purity
of the encoding, the order of Bob's measurement operations or
measurement symmetry is needed. Public randomness does not weaken
the exact bounds, since the finite input and matching sets permit
fixing a seed that is correct simultaneously on every input pair.
The complete operator inequalities are given in
Appendix~\ref{app:quantum}. Together with the phase protocol, the
bound proves the optimal leading logarithmic qubit cost, without
requiring the phase-code dimension to be exactly minimal.

The separation concerns the optimal costs of the two tasks, but
it also raises a question about the information carried by the
exclusion message itself. For uniform inputs, the phase ensemble
has average state $I/n$, whereas the conditional state given
$\Delta_e=b$ satisfies $\bar\rho_{e,b}\le2I/n$. This limits the
retrieval probability of any decoder, as formalized below together
with the orthogonality criterion for phase states
(Appendix~\ref{app:phase}).

\begin{corollary}[Phase-code retrieval and orthogonality]\label{cor:phase}
For uniform $x$, any nonzero decoding effect $G$, fixed edge $e$, and label $b$ satisfy $\Pr(\Delta_e=b\mid G)\le2/p$. The bound is attainable. Two phase states are orthogonal if and only if each field symbol occurs equally often in their componentwise differences.
\end{corollary}
Thus the same phase message permits certain exclusion but retrieves
a matched difference with optimal probability only $2/p$.
Table~\ref{tab:communication} summarizes the communication costs
and the phase-code performance.

\begin{table}[tb]
\caption{\footnotesize Unassisted one-way communication
without parity obliviousness, for fixed odd prime $p$.
Costs are classical bits or quantum qubits. Fixed
advantage $\eta$ denotes a constant improvement over
the random-guessing baseline, strictly below perfect
success. The last row gives the optimal success
probability for the phase encoding.}
\label{tab:communication}
\centering\footnotesize
\setlength{\tabcolsep}{1.5pt}
\renewcommand{\arraystretch}{1.25}
\begin{ruledtabular}
\begin{tabular}{@{}lccc@{}}
& \shortstack{Binary\\HM / exclusion}
& \shortstack{Odd-prime\\DHM}
& \shortstack{Odd-prime\\DHAM}\\\hline
\shortstack[l]{Exact classical\\communication}
&$\Theta(n)$
&$\Theta_p(n)$
&$\Theta_p(n)$\\
\shortstack[l]{Exact quantum\\communication}
&$\Theta(\log n)$
&$\Theta_p(n)$
&$\log_2 n+O_p(1)$\\
\shortstack[l]{Classical\\fixed advantage}
&$\Theta_{\eta}(\sqrt n)$
&$\Theta_{p,\eta}(\sqrt n)$
&$\Theta_{p,\eta}(\sqrt n)$\\
\shortstack[l]{Optimal phase-code\\success}
&$1$
&$2/p$
&$1$\\
\end{tabular}
\end{ruledtabular}
\end{table}

\textit{Preparation contextuality and noncontextual bounds.}---
The communication separation does not rely on any restriction on
Alice's preparations. We now impose an information restriction to
ask a different question: can the perfect exclusion statistics of
the phase protocol admit a preparation-noncontextual explanation?
Equation~\eqref{eq:em-coherence} provides the natural starting point.
Each off-diagonal entry of the phase-state density operator depends
only on a pair difference. We formalize this restriction as
\emph{pair-difference parity obliviousness}.

With $\chi_r(x)=\omega^{r\cdot x}$, the allowed nonconstant
Fourier directions and their forbidden complement are
\begin{equation}
\begin{gathered}
\mathcal D_{n,p}=\{a(\mathbf e_j-\mathbf e_i):
i<j,\ a\in\Fp^*\},\\
\mathcal F_{n,p}=\Fp^n\setminus
(\{0\}\cup\mathcal D_{n,p}).
\end{gathered}
\label{eq:forbidden}
\end{equation}
For every forbidden $r$, preparing $x$ uniformly from the
class $r\cdot x=k$ must be operationally indistinguishable
from preparing it uniformly from the class $r\cdot x=k'$.
In quantum theory, this requires
\begin{equation}
\overline\rho_{r,k}
=p^{1-n}\!\sum_{x:r\cdot x=k}\rho_x
=\overline\rho_{r,k'}
\quad(k,k'\in\Fp).
\label{eq:equivs}
\end{equation}
When public randomness is present, these operational equivalences
are imposed on the complete message available to Bob, including
the seed. Our phase ensemble satisfies
$\overline\rho_{r,k}=I/n$ for every forbidden $r$,
so its perfect exclusion score survives the constraint.
For $p=2$, these reduce to the familiar pair-parity
equivalences of PoHM~\cite{pohm}. We denote the
restricted retrieval and exclusion tasks by PoDHM
and PoDHAM, respectively. The corresponding
restrictions for binary PoHM and odd-prime PoDHAM
are illustrated in Fig.~\ref{fig:tasks}(c) and (d).

Since the quantum protocol remains exact under these equivalences,
we next characterize the corresponding noncontextual strategies.
Suppose that each preparation $x$ produces an underlying
physical state $\lambda\in\Lambda$, described by a probability
measure $\mu_x$. Bob's response to a matching may depend on
$\lambda$, but operationally indistinguishable mixtures must
induce the same distribution over $\lambda$
~\cite{spekkens,SchmidEquv2018}.
Let $\nu=p^{-n}\sum_x\mu_x$ and define
$f_\lambda(x)=d\mu_x/d\nu(\lambda)$, with the derivative
taken on the support of $\nu$. For uniformly sampled inputs,
$p^{-n}f_\lambda(x)$ is Alice's posterior probability
of having prepared $x$, conditioned on $\lambda$;
in particular, $f_\lambda\ge0$ and
$\mathbb E_x f_\lambda(x)=1$.
Preparation noncontextuality imposes the operational
equivalences of Eq.~\eqref{eq:equivs} at the ontic level,
eliminating every forbidden Fourier component.
Consequently, for $\nu$-almost every $\lambda$,
\begin{equation}
\begin{gathered}
f_\lambda(x)=1+
\sum_{i<j}h_{ij,\lambda}(x_j-x_i)\ge0,\\
\sum_{b\in\Fp}h_{ij,\lambda}(b)=0.
\end{gathered}
\label{eq:posterior}
\end{equation}
Thus, preparation-noncontextual explanations reduce to
nonnegative probability distributions whose only
nonconstant Fourier components are pair-difference
terms. This reduction accommodates arbitrary measurable
ontic spaces and assumes neither outcome determinism
nor measurement noncontextuality
(Appendix~\ref{app:ontic}).

For $e=(i,j)$, the probability assigned to a difference
$b$ at the ontic state is
$q_\lambda(e,b)=\Pr(\Delta_e=b\mid\lambda)
=[1+h_{ij,\lambda}(b)]/p$.
Bob can claim to exclude $b$ conclusively only if this
probability is zero. More generally, for a specified matching
his best exclusion success is
\begin{equation}
S_{\excl}(f_\lambda,M)
=1-\min_{e\in M,\,b\in\Fp}q_\lambda(e,b).
\label{eq:ontic-score}
\end{equation}
Optimizing and averaging this expression over permitted
posteriors yields the preparation-noncontextual bound.
Conversely, translation averaging implements any normalized,
nonnegative posterior of the form in Eq.~\eqref{eq:posterior}
as a parity-oblivious classical strategy. Hence tightness
claims below refer to operationally realizable models,
not merely formal Fourier functions (see Appendix~\ref{app:ontic}).

The decisive constraint for perfect exclusion is therefore the
positivity of a common posterior across different matchings.
\begin{theorem}[Triangle obstruction]\label{thm:triangle}
For any prime $p$, all edges on which an allowed ontic posterior conclusively excludes a value lie within a single triangle. Thus, for every even $n\ge6$, no preparation-noncontextual model succeeds perfectly over all perfect matchings, whereas the phase protocol does. At $n=4$ a perfect parity-oblivious classical strategy exists.
\end{theorem}
For disjoint edges, excluding two values would make the
conditional average of $f_\lambda$ equal to $-1$, contradicting
nonnegativity. Positivity also rules out three conclusively
excluded edges sharing a vertex. A graph containing neither two
disjoint edges nor a three-edge star must lie within a triangle.
For $n\ge6$, some perfect matching avoids that triangle, ruling
out perfect noncontextual exclusion over the full matching family
(Appendix~\ref{app:triangle}).

This geometric obstruction suggests a way to reduce the number
of measurement settings. Rather than use all perfect matchings,
we seek the smallest family that no triangle intersects in every
setting. The following result gives both the exact criterion and
a tight noncontextual score for a minimal construction.

\begin{figure}[tb]
\centering
\includegraphics[width=0.9\columnwidth]{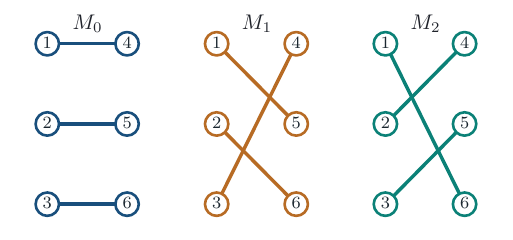}
\caption{\footnotesize Three perfect matchings at $n=6$: $M_0=\{14,25,36\}$, $M_1=\{15,26,34\}$ and $M_2=\{16,24,35\}$. Their union is $K_{3,3}$, so no triangle meets all three. This gives the minimal AVN witness, with $P_{\NC}^{(3)}=1-1/p^2$ for odd primes.}
\label{fig:settings}
\end{figure}

\begin{theorem}[Matching families and three-setting bound]\label{thm:settings}
A matching family $\mathcal S\subseteq\Mn$ admits perfect parity-oblivious classical, equivalently preparation-noncontextual, exclusion if and only if a triangle intersects every $M\in\mathcal S$. For even $n\ge6$, three settings are necessary and sufficient for AVN contextuality. With disjoint vertex sets $\{a_r\}$, $\{b_r\}$, $n=2m$, take
\begin{equation}
M_s=\{\{a_r,b_{r+s}\}:r\in\mathbb Z_m\},\quad s=0,1,2.
\label{eq:settings}
\end{equation}
For these three cyclic settings, the optimal uniform noncontextual exclusion score is
\begin{equation}
P_{\NC}^{(3)}(n,p)=1-\min\left\{\frac1{3p},\frac1{p^2}\right\}.
\label{eq:3bound}
\end{equation}
For odd primes this gives $1-1/p^2$ against $P_Q^{(3)}=1$.
At $p=2$, Eq.~\eqref{eq:3bound} recovers the previously
established $5/6$ score~\cite{pohm}; the binary triangle
restriction is recovered by Theorem~\ref{thm:triangle}.
\end{theorem}
Every pair of matchings has a common triangle transversal, so
two settings cannot reveal the contradiction. The three cyclic
matchings in Eq.~\eqref{eq:settings}, however, have a triangle-free
bipartite union, as illustrated for $n=6$ in Fig.~\ref{fig:settings}.
Nonnegative triangle and star posteriors attain the bound in
Eq.~\eqref{eq:3bound} under the stipulated operational equivalences
(Appendix~\ref{app:three}).

The tightness of this bound depends on which equivalences are
imposed. In particular, Eq.~\eqref{eq:em-coherence} shows that
the characterized phase ensemble has only the first positive and
negative harmonics on each edge line. For $p\ge5$, this is a
strict restriction compared with Eq.~\eqref{eq:posterior}, which
allows all $p-1$ nonzero harmonics. Imposing all preparation-mixture
equivalences of the characterized phase ensemble therefore leads
to the stronger upper bound
\begin{equation}
P_{\NC,\mathrm{phase}}^{(3)}\le1-\frac1{3p},\qquad p\ge5.
\label{eq:phase-nc}
\end{equation}
Since this is strictly below $1-1/p^2$, the stipulated benchmark
cannot be saturated under the additional equivalences. We do not
claim that Eq.~\eqref{eq:phase-nc} is tight; its derivation is
provided in Appendix~\ref{app:three}.

Besides revealing contextuality with only three settings, the
minimal matching family retains a separation between the exact
communication dimensions of retrieval and exclusion. At $n=6$,
exact DHM retrieval for the matchings in
Fig.~\ref{fig:settings} requires Hilbert-space dimension
$d_{\mathrm{DHM}}\ge p^2$, whereas perfect DHAM exclusion
is achievable in dimension six. Hence
$d_{\mathrm{DHM}}\ge p^2>6\ge d_{\mathrm{DHAM}}$
for every odd prime $p$ (Appendix~\ref{app:three}).

Having identified the minimal witness, we return to the full
family of perfect matchings to determine how much a noncontextual
strategy can gain over random guessing as the input length grows.
\begin{theorem}[Full-family noncontextual scaling]\label{thm:full}
For fixed prime $p$, uniformly distributed perfect matchings satisfy
\begin{equation}
\begin{aligned}
P_{\NC}^{\excl}(n,p)&=1-\frac1p+\Theta_p(n^{-1/2}),\\
P_{\NC}^{\ret}(n,p)&=\frac1p+\Theta_p(n^{-1/2}).
\end{aligned}
\label{eq:fullbound}
\end{equation}
For even $n\ge6$, $P_{\NC}^{\excl}\le1-\min\{1/(3p),1/p^2\}$.
\end{theorem}
Hypercontractivity bounds the total edge-dependent Fourier
weight of a nonnegative posterior, while explicit positive
posteriors attain the $n^{-1/2}$ advantage
(Appendix~\ref{app:full}). Thus increasing the number of input
positions suppresses the noncontextual advantage over guessing,
whereas the quantum exclusion score remains unity. The privacy
constraint differs from that of coordinate-wise retrieval and
exclusion games, as summarized in Table~\ref{tab:tasks}.

\begin{table}[tb]
\caption{\footnotesize Comparison of parity-oblivious
communication tasks. The benchmarks correspond to
their respective operational equivalences and
communication scenarios.}
\label{tab:tasks}
\centering\scriptsize
\setlength{\tabcolsep}{2.5pt}
\renewcommand{\arraystretch}{1.17}
\begin{ruledtabular}
\begin{tabular}{@{}p{0.16\columnwidth}
p{0.39\columnwidth}p{0.37\columnwidth}@{}}

\textbf{Task}&
\textbf{Objective and privacy}&
\textbf{Benchmark}\\\hline

PORAC~\cite{porac,hameedi}&
Retrieve $x_y$ for specified $y$;
only single-coordinate characters exempt.&
$P_{\NC}=(n+p-1)/(np)$;
quantum advantage depends on message dimension.\\

POREC~\cite{porec}&
Exclude $x_y$ for specified $y$;
only single-coordinate characters exempt.&
Two ternary digits:
$P_{\NC}=5/6$,
$P_Q^{d=2}=2/3+1/(3\sqrt2)$.\\

Binary PoHM~\cite{pohm}&
Retrieve $x_i\oplus x_j$ on a chosen matched edge;
only weight-two parities exempt.&
$n=6$, three cyclic settings:
$P_{\NC}=5/6$, $P_Q=1$.\\

Odd-prime PoDHAM (this work)&
Exclude $\Delta_{ij}(x)$ on a chosen matched edge;
only pair-difference characters exempt.&
Three cyclic settings of Eq.~\eqref{eq:settings},
even $n\ge6$:
$P_{\NC}=1-1/p^2$, $P_Q=1$.\\

\end{tabular}
\end{ruledtabular}
\end{table}

\textit{Measurement certification.}---
The preceding contextuality results follow from restrictions on
preparation mixtures without assuming a particular measurement
realization. We now characterize the phase source and ask what
the same exclusion zeros imply about Bob's measurement.
In particular, can projective measurements, supplemented by
classical randomness and postprocessing, reproduce the exclusion
statistics? Generalized qubit POVMs have been implemented
experimentally~\cite{Zhao2015} and studied in characterized
prepare-and-measure certification scenarios~\cite{guerini,oszmaniec,tavakoli,Diebra2026,YaoWang2026}.

\begin{proposition}[Conclusive effects and simulability]\label{prop:measurement}
For the characterized phase preparations, every nonzero effect $F_{e,b}$ excluding $b$ whenever $\Delta_e=b$ satisfies
\begin{equation}
\supp F_{e,b}\subseteq\operatorname{span}\{\ket{\chi_{e,b}}\}.
\label{eq:rigidity}
\end{equation}
For odd $p$ perfect edge-resolved exclusion needs at least three nonzero qubit effects; for $p=3$ the label-aggregated effects uniquely form the antitrine POVM. For odd $p$, uniform inputs, a characterized pure phase source,
and edge-resolved qubit readout, projective-simulable
measurements satisfy the tight bound
\begin{equation}
S_{\PS}(p)=1-\frac{1-\cos(\pi/2p)}p.
\label{eq:ps}
\end{equation}
\end{proposition}
The support restriction follows because the exclusion probability
must vanish on every preparation having the reported difference.
To obtain the projective-simulability bound, condition on the
projection onto a fixed edge and define
$\sigma_b=\ket{\phi_b}\bra{\phi_b}$, with
$\ket{\phi_b}=(\ket i+\omega^b\ket j)/\sqrt2$.
Since the exclusion score is linear in the readout, the optimal
projective-simulable strategy reduces to a binary projective
measurement with two distinct exclusion labels $a,b$.
Its minimum failure probability is
$[1-\|\sigma_a-\sigma_b\|_1/2]/p$.
Maximizing the trace distance over the labels yields
Eq.~\eqref{eq:ps}; the full optimization and the support
argument are given in Appendix~\ref{app:measurement}.
For $p=3$,
\begin{equation}
\frac89=P_{\NC}^{(3)}<\frac{4+\sqrt3}6=S_{\PS}(3)<1.
\label{eq:thresh}
\end{equation}
Although both bounds concern the same ternary exclusion score,
they certify different resources. The noncontextual bound relies
on operational equivalences~\cite{spekkens,SchmidEquv2018}, whereas
the projective-simulability bound assumes a characterized source
and edge projection and is not device independent
~\cite{guerini,oszmaniec,tavakoli}.

\begin{figure}[tb]
\centering
\includegraphics[width=0.9\columnwidth]{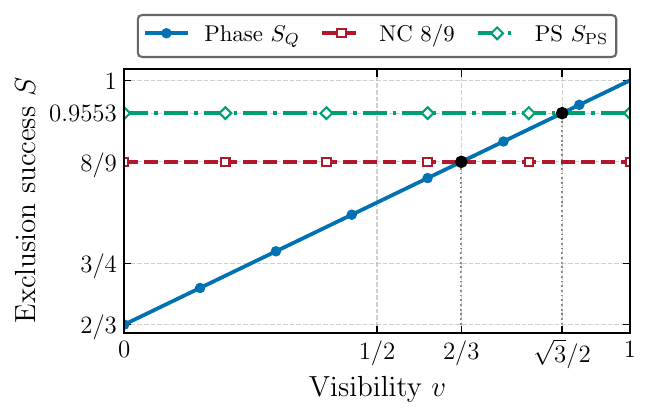}
\caption{\footnotesize Ternary exclusion score $S_Q(v)=(2+v)/3$, stipulated preparation-noncontextual bound $8/9$, and characterized \emph{pure-source} projective-simulability reference $(4+\sqrt3)/6$. The intersections at $v=2/3$ and $\sqrt3/2$ pertain to distinct certification models.}
\label{fig:thresholds}
\end{figure}

To compare their robustness, consider the depolarized phase
preparations $\rho_x(v)=v\ket{\psi_x}\bra{\psi_x}+(1-v)I/n$.
With the ideal exclusion POVM, the score becomes

\begin{equation}
S_Q(v)=1-\frac{1-v}p.
\label{eq:noise}
\end{equation}
For $p=3$, Eq.~\eqref{eq:noise} exceeds the noncontextual bound
when $v>2/3$. It exceeds the fixed pure-source projective-simulability
reference when $v>\sqrt3/2$, as shown in Fig.~\ref{fig:thresholds}.
The latter comparison is conservative: if the noisy source itself
is characterized, the projective-simulability benchmark must be
optimized for the actual conditional preparations
$\sigma_b(v)=v\sigma_b+(1-v)I_2/2$. Their pairwise trace distances
are reduced by a factor $v$, giving the tight bound
\begin{equation}
S_{\PS}(p,v)=1-\frac{1-v\cos(\pi/2p)}p.
\label{eq:em-noisyps}
\end{equation}
Under these calibrated-source assumptions, the ideal ternary
antitrine readout exceeds Eq.~\eqref{eq:em-noisyps} for every
$v>0$. The fixed-reference comparison in Fig.~\ref{fig:thresholds}
should therefore not be interpreted as the optimized noisy-source
threshold. Appendix~\ref{app:measurement} derives quantitative
resource-weight bounds for both certification models, illustrated
in Fig.~\ref{fig:resource_weights}.

\textit{Experimental perspective.}---
The minimal ternary witness offers a common experimental platform
for the contextuality and measurement-certification tests, although
their characterization requirements differ. For $(n,p)=(6,3)$,
the phase protocol uses a
six-mode phase source and three matching measurements, each
with nine edge-resolved outcomes. The phases
$\{0,2\pi/3,4\pi/3\}$ encode $3^5=243$ distinct rays after
global shifts are identified, as in Eq.~\eqref{eq:em-coherence}.
Programmable six-mode photonics and optical implementations of
three-outcome qubit POVMs provide relevant experimental building
blocks~\cite{Giordani2023,Zhao2015}.

For contextuality certification, the principal overhead lies
in establishing the preparation-mixture equivalences rather than
implementing the three witness measurements.
There are $349$ forbidden Fourier lines in the full input
space, of which $106$ remain after quotienting global shifts.
Each forbidden quotient line specifies three uniform mixtures
of $81$ rays that ideally equal $I_6/6$. The three ideal
matching measurements provide only $22$ independent
probabilities per normalized preparation, fewer than the $35$
required for unrestricted six-mode tomography. Thus the three
witness settings alone cannot establish the requisite
operational equivalences. Auxiliary fiducial measurements,
justified source calibration, or verified secondary preparations
are necessary to address these equivalences~\cite{mazurek,Pusey2018}.
Approximate mixture equality does not automatically justify the
ideal $8/9$ preparation-noncontextual bound. By contrast, the measurement-certification test relies on
characterization of the phase preparations and edge-resolved
readout~\cite{Brinster2026}. Appendix~\ref{app:implementation}
details the preparation count, measurement completeness, and
operational requirements.

\textit{Discussion.}---Our results establish a fundamental distinction
between retrieving relational information and excluding
one of its possible values. Quantum interference enables
perfect exclusion without identifying the correct matched
difference, and encoding-independent bounds show that
the exponential separation is intrinsic to the tasks,
not to a particular encoding. Unlike parity-oblivious
random access and exclusion codes~\cite{porec}, this
separation arises without preparation restrictions.

The foundational significance of conclusive exclusion
lies in the consistency of its zero-probability
predictions. Although quantum measurements can exclude
a value on every matched edge, a preparation-noncontextual
description must reproduce these exclusions through
a common positive ontic posterior, whose support is
subject to stringent geometric constraints.
The resulting bounds also depend on which operational
equivalences are imposed, as illustrated by the
stronger restriction for the characterized phase
ensemble. From a different perspective, the same
exclusion zeros constrain the structure of Bob's
measurement, revealing qubit effects beyond projective
simulability when preparations are characterized and
outcomes are resolved edge by edge. Thus conclusive
exclusion probes both the limitations of noncontextual
explanations and the structure of quantum measurements,
under distinct operational assumptions.

\textit{Open problems.}---
For fixed $2/p<s<1$, does retrieving a matched
difference with success probability at least $s$ require
superlogarithmic quantum communication? The phase
protocol attains $2/p$ with $\lceil\log_2 n\rceil$ qubits,
while public sampling gives an $O_{p,s}(\sqrt n)$-qubit
upper bound (Appendix~\ref{app:classical}, part (c)). Other challenges include determining
the exact exclusion dimension within
$1+(n-1)/(p-1)\le d_{\min}\le n$ and understanding
preshared-entanglement assistance~\cite{chakraborty}.
Extending the cyclotomic orthogonality criterion
and field-based contextuality theorems to composite
symbols, sequential exclusion
and instrument certification~\cite{Mohan2019,Foletto2020,
Roy2026Sequential,Schmid2026PTM}, and imperfect operational
equivalences~\cite{mazurek,Pusey2018,Giordani2023}
may enable cryptographic, streaming, and
semi-device-independent applications~\cite{brask,lunghi}.
More broadly, conclusive exclusion emerges as a
resource for quantum communication savings and
operational certification of nonclassicality.

\textit{Acknowledgment.}--- P. R. acknowledges financial support from SNBNCBS, Kolkata.

\textbf{Declaration of AI use:} All scientific ideas and results are the author's own. 
ChatGPT (5.6 Sol, OpenAI) was used to assist with language 
refinement, \LaTeX{} formatting, bibliographic verification, 
and manuscript consistency checks. The author has carefully 
reviewed the manuscript and assumes full responsibility for 
its content, accuracy, and scientific integrity.

%

\onecolumngrid
\appendix
\setcounter{secnumdepth}{2}
\numberwithin{lemma}{section}
\counterwithin{figure}{section}
\renewcommand{\thefigure}{\thesection\arabic{figure}}
\renewcommand{\theHfigure}{appendix.\Alph{section}.\arabic{figure}}

\section{Proof of Theorem~\ref{thm:classical}: classical communication bounds}\label{app:classical}

\noindent\textbf{Theorem~\ref{thm:classical} (Classical communication).}
For every fixed prime $p$,
\begin{equation*}
C_0^\rightarrow(\mathrm{DHM}_{n,p})=\Theta_p(n),\qquad
C_0^\rightarrow(\mathrm{DHAM}_{n,p})=\Theta_p(n).
\end{equation*}
For any fixed positive advantage above the respective random-guessing
probabilities and strictly below perfect success, both public-coin
costs are $\Theta_{p,\eta}(\sqrt n)$.

\smallskip\noindent\textit{Proof.}---\enspace
We treat the zero-error and fixed-advantage regimes separately.

\smallskip\noindent\textit{(a) Exact classical communication:}\enspace
Fix a deterministic zero-error message $z$ and let
$A\subseteq\Fp^n$ be the set of inputs producing it. Because Bob
receives no further information about which $x\in A$ was sent, any
output made after receiving $z$ must be valid for \emph{every} input
in $A$. Define a graph on $[n]$ by
\begin{equation}
G_A=\bigl\{e:\text{there exists }b_e\in\Fp
\text{ with }\Delta_e(x)\ne b_e\text{ for all }x\in A\bigr\}.
\label{eq:supp-classical-graph}
\end{equation}
An edge belongs to $G_A$ precisely when Bob can exclude at least one
difference there using only the message $z$. Exact DHAM is possible for
all matchings only if $G_A$ intersects every perfect matching.

Suppose every connected component of $G_A$ had at most $n/2$
vertices. Joining vertices from different components creates a
complete multipartite graph with a perfect matching, because a complete multipartite graph on an even vertex set has
a perfect matching whenever no part contains more than half the
vertices: repeatedly match vertices from the two largest nonempty
parts. Such a matching has no edge in $G_A$, a contradiction. Thus $G_A$ contains
a component of size $m>n/2$. Choose a spanning tree of that component
and orient it away from a root. There are at most $p$ possible
values at the root. On every tree edge, the defining exclusion
constraint forbids one difference; after the parent value is fixed,
the child therefore has at most $p-1$ values. The remaining $n-m$
coordinates are unrestricted. Consequently,
\begin{equation}
|A|\le p^{n-m}p(p-1)^{m-1}
\le p^n\left(\frac{p-1}{p}\right)^{n/2}.
\label{eq:supp-cell-bound}
\end{equation}
At least $p^n/\max_A|A|\ge[p/(p-1)]^{n/2}$ distinct messages are
necessary. Taking logarithms proves an $\Omega_p(n)$-bit exclusion
lower bound. An exact DHM decoder also yields an exact DHAM decoder by
replacing its correct label by any different one; thus the same
lower bound applies to retrieval.

For the upper bound, publicly fix any set of $n/2+1$ positions. Alice
sends their values relative to one reference position. Every perfect
matching contains an edge with both endpoints in this set, since
there are more selected vertices than matching edges. Bob can compute
its difference exactly, and either report it (DHM) or report another
value (DHAM). The message uses $O_p(n)$ bits.

At finite $n$ the two exact costs can differ. For $n=4,p=5$, Alice can send a label from $\{0,1,2,3\}$ absent among the three differences on a fixed triangle. Two bits suffice for exclusion because every matching meets that triangle. For retrieval, two distinct settings fix differences on two distinct edges in each deterministic message cell. These independent constraints leave at most $5^2$ inputs per cell, so at least $25$ messages are necessary. Sending the two relative values on the triangle attains this count, giving five bits in the fixed-length convention.

\smallskip\noindent\textit{(b) Fixed-advantage lower bound:}\enspace
We use uniform inputs and a uniformly random perfect matching. This
suffices for a public-coin worst-case lower bound: a protocol whose
success is guaranteed on every input has at least the same average
success, and some fixing of its public coins attains that average.

For a deterministic message cell $A$ of uniform-input probability
$\alpha=|A|/p^n>0$, put $f=\ind_A$ and $g=f/\alpha$.
We normalize Fourier coefficients by
\begin{equation}
\widehat f(r)=\E_{x\in\Fp^n}f(x)\omega^{-r\cdot x}.
\end{equation}
The finite-field hypercontractive inequality (Lemma 4 of
Ref.~\cite{guruswami}) states that
\begin{equation}
\sum_{r\in\Fp^n}\delta^{|r|}|\widehat f(r)|^2
\le\alpha^{2/(1+p\delta)},
\qquad 0\le\delta\le1/p,
\label{eq:supp-fourier-inequality}
\end{equation}
where $|r|$ is the number of nonzero coordinates of $r$.
Set $L=\ln(1/\alpha)$ and
$\delta=[p\max\{1,L\}]^{-1}$. Keeping only the Fourier terms of
weight two in Eq.~\eqref{eq:supp-fourier-inequality} and dividing by
$\alpha^2$ gives
\begin{equation}
\sum_{|r|=2}|\widehat g(r)|^2
\le\mathrm e^2p^2(1+L)^2.
\label{eq:supp-level-two}
\end{equation}
Here the factor $\mathrm e^2$ bounds the residual term
$\alpha^{2/(1+p\delta)-2}$ after the above choice of $\delta$.

For an edge $e$, write
\begin{equation}
r_e(b)=\Pr(\Delta_e=b\mid A),\quad
u_e(b)=r_e(b)-\frac1p,
\end{equation}
and define its best exclusion and retrieval bonuses as
\begin{equation}
\delta_e=-\min_bu_e(b),\qquad
\gamma_e=\max_bu_e(b).
\label{eq:supp-bonuses}
\end{equation}
Because $\sum_bu_e(b)=0$, either extreme
$a_e\in\{\delta_e,\gamma_e\}$ satisfies
\begin{equation}
\sum_{b\in\Fp}u_e(b)^2\ge\frac p{p-1}a_e^2.
\label{eq:supp-extreme}
\end{equation}
Indeed, once one entry has magnitude $a_e$, the other $p-1$
entries have total $\mp a_e$, and Cauchy--Schwarz minimizes their
squared sum when they are equal.
Fourier inversion along the one-dimensional line associated with
$e=(i,j)$ and Parseval's identity give
\begin{equation}
p\sum_bu_e(b)^2
=\sum_{a\in\Fp^*}
\left|\widehat g\bigl(a(\mathbf e_j-\mathbf e_i)\bigr)\right|^2.
\end{equation}
The Fourier lines associated with different edges are disjoint, and
all their nonzero vectors have Hamming weight two. Combining this
identity with Eqs.~\eqref{eq:supp-level-two} and
\eqref{eq:supp-extreme} yields
\begin{equation}
\sum_e\delta_e^2,\quad\sum_e\gamma_e^2
\le\mathrm e^2(p-1)(1+L)^2.
\label{eq:supp-classical-l2}
\end{equation}

On a fixed matching, Bob's best advantage is the largest bonus
among its edges. Each edge occurs in a uniformly random matching
with probability $1/(n-1)$. Applying Cauchy--Schwarz and then
Eq.~\eqref{eq:supp-classical-l2} gives, for either kind of bonus,
\begin{equation}
\E_M\max_{e\in M}\delta_e,\quad
\E_M\max_{e\in M}\gamma_e
\le\frac{\mathrm e\sqrt{p-1}}{\sqrt{n-1}}(1+L).
\label{eq:supp-matching-average}
\end{equation}
Now average over the message cells with probabilities $\alpha_z$.
For a message of at most $c$ bits, its entropy obeys
$\sum_z\alpha_z\ln(1/\alpha_z)\le c\ln2$.
Thus the overall success probabilities satisfy the explicit bounds
\begin{align}
P_C^{\excl}&\le1-\frac1p+
\frac{\mathrm e\sqrt{p-1}}{\sqrt{n-1}}(1+c\ln2),
\label{eq:classical-explicit-ex}\\
P_C^{\ret}&\le\frac1p+
\frac{\mathrm e\sqrt{p-1}}{\sqrt{n-1}}(1+c\ln2).
\label{eq:classical-explicit-ret}
\end{align}
A fixed positive advantage therefore requires $c=\Omega_{p,\eta}
(\sqrt n)$. The distributional-to-public-coin step is the standard
fixing-of-randomness argument~\cite{yao,kn}.

\smallskip\noindent\textit{(c) Matching public-coin protocol:}\enspace
Publicly sample $k\le n/2$ positions uniformly and have Alice send
their values relative to one sampled position. Bob succeeds exactly
whenever the sample contains both endpoints of some edge of his
matching. To avoid containing a complete edge, the sample must
choose $k$ different matching edges and one endpoint from each.
Hence the failure probability of finding an internal edge is
\begin{equation}
q(n,k)=\frac{2^k\binom{n/2}{k}}{\binom nk}
=\prod_{j=1}^k\frac{n-2j+2}{n-j+1}.
\label{eq:sample-noedge}
\end{equation}
If no such edge is sampled, Bob outputs a uniformly random label.
His total success probabilities are therefore
\begin{align}
P_{\excl}&=1-\frac1p+\frac{1-q(n,k)}p,\\
P_{\ret}&=\frac1p+\left(1-\frac1p\right)[1-q(n,k)].
\end{align}
Choosing $k=\beta\sqrt n+O(1)$ for any fixed $\beta>0$ gives
$q(n,k)\to\exp(-\beta^2/2)$. By choosing $\beta$ according to the
desired advantage, both protocols use $O_{p,\eta}(\sqrt n)$ bits.
This completes the proof of Theorem~\ref{thm:classical}. \hfill$\square$
\section{Proof of Theorem~\ref{thm:quantum}: exact quantum communication separation}\label{app:quantum}

\noindent\textbf{Theorem~\ref{thm:quantum} (Exact quantum separation).}
For every fixed odd prime $p$ and even $n\ge4$,
the exact unassisted one-way quantum communication costs satisfy
\begin{equation*}
\begin{aligned}
Q_0^\rightarrow(\mathrm{DHM}_{n,p})&=\Theta_p(n),\\
Q_0^\rightarrow(\mathrm{DHAM}_{n,p})&=\log_2 n+O_p(1).
\end{aligned}
\end{equation*}
We establish the encoding-independent finite-size bounds of
Eqs.~\eqref{eq:qlower} and \eqref{eq:qudimension} of the main text,
including the explicit $n$-dimensional exclusion construction.
All converses allow arbitrary mixed-state encodings and POVMs,
without preshared entanglement.

\smallskip\noindent\textit{Proof.}---\enspace
We establish the upper and lower bounds, including the encoding-independent exclusion dimension lower bound. The exclusion protocol is explicit;
the retrieval lower bound applies to arbitrary, possibly mixed-state
encodings and arbitrary measurements.

\smallskip\noindent\textit{(a) Exact exclusion with logarithmically many qubits:}\enspace
Alice prepares the phase state in Eq.~\eqref{eq:phase} of the main text.
For a matched edge $e=(i,j)$, Bob uses the effects $E_{e,b}$ in
Eq.~\eqref{eq:povm} of the main text. Their normalization follows from
$\sum_{b\in\Fp}\omega^b=0$:
\begin{equation}
\sum_{b\in\Fp}E_{e,b}
=\frac2p\sum_{b\in\Fp}\ket{\chi_{e,b}}\bra{\chi_{e,b}}
=\ket i\bra i+\ket j\bra j.
\end{equation}
Since the edges of a perfect matching partition $[n]$, summing these
effects over $e\in M$ gives the identity on the full message space.
Thus the collection $\{E_{e,b}:e\in M,\ b\in\Fp\}$ is a POVM for every
matching available to Bob. Its outcome distribution is
\begin{equation}
\langle\psi_x|E_{e,b}|\psi_x\rangle
=\frac1{np}\bigl|1-\omega^{\Delta_e(x)-b}\bigr|^2.
\end{equation}
For the correct difference $b=\Delta_e(x)$ this probability is exactly
zero. Every outcome therefore reports an incorrect difference, as DHAM
requires, and no inconclusive outcome is needed. The transmitted state
has dimension $n$, attaining the upper bound in
Eq.~\eqref{eq:qudimension} of the main text.

\smallskip\noindent\textit{(b) Encoding-independent lower bound for exact retrieval:}\enspace
We first identify pairs of inputs that no exact DHM protocol can encode
into overlapping quantum supports. For $z\in\Fp^n$, write
$N_a(z)=|\{i:z_i=a\}|$. If
\begin{equation}
\max_{a\in\Fp}N_a(z)\le\frac n2,
\label{eq:supp-balanced}
\end{equation}
there exists a perfect matching that pairs only positions carrying
different symbols of $z$. To see this, group the positions by their
symbol. The graph joining positions in different groups is complete
multipartite. When no group contains more than half the vertices, it
has a perfect matching: match vertices from two largest nonempty
groups, remove them, and repeat. The same size condition holds at each
step, giving the claimed matching by induction.

Now set $z=y-x$. On an edge $e=(i,j)$ that connects different symbols of
$z$, one has
\begin{equation}
\Delta_e(y)-\Delta_e(x)=z_j-z_i\ne0.
\end{equation}
Consequently, if $z$ obeys Eq.~\eqref{eq:supp-balanced}, there is a
matching for which no valid DHM output $(e,b)$ for $x$ is also valid for
$y$. We call such inputs \emph{incompatible}.

It remains to construct exponentially many pairwise incompatible
inputs. Let
\begin{equation}
\mathcal B=\{z\in\Fp^n:\max_aN_a(z)>n/2\}
\end{equation}
be the set of unbalanced differences. For uniformly random $z$,
each count $N_a(z)$ has the binomial distribution
$\mathrm{Bin}(n,1/p)$. A union bound together with the binomial
large-deviation inequality yields
\begin{align}
\frac{|\mathcal B|}{p^n}
&\le p\exp(-nD_p),\\
D_p&=D\!\left(\frac12\middle\|\frac1p\right)
=\ln\frac{p}{2\sqrt{p-1}}>0,
\label{eq:supp-rate}
\end{align}
where positivity uses $p>2$. Form a graph on $\Fp^n$, connecting
distinct inputs $x,y$ whenever $y-x\in\mathcal B$. Every vertex has
degree at most $|\mathcal B|$. Greedily selecting a vertex and deleting
its neighbors therefore gives an independent set $\mathcal C$ with
\begin{equation}
|\mathcal C|\ge\frac{p^n}{|\mathcal B|+1}
\ge\frac{e^{nD_p}}{2p}.
\label{eq:supp-code}
\end{equation}
For every distinct $x,y\in\mathcal C$, their difference is balanced,
so they are incompatible in the above sense.

Consider any zero-error DHM encoding, with density operator $\rho_x$
for input $x$. For incompatible $x,y$, choose a matching that makes
their valid output sets disjoint. Sum Bob's effects for outputs valid
for $x$ into a single effect $F$. Exactness gives
\begin{equation}
0\le F\le I,\qquad
\Tr(F\rho_x)=1,\qquad \Tr(F\rho_y)=0.
\label{eq:supp-perfect-discrimination}
\end{equation}
Because $F$ is bounded between zero and the identity, the first trace
condition places $\supp\rho_x$ in its eigenvalue-one subspace; the
second places $\supp\rho_y$ in its kernel. These subspaces are
orthogonal. Thus all $|\mathcal C|$ messages have pairwise orthogonal
supports, irrespective of their purity or the decoder used. Their
Hilbert space must have dimension at least $|\mathcal C|$.

Taking base-two logarithms in Eq.~\eqref{eq:supp-code} yields
\begin{equation}
\log_2\dim\mathcal H
\ge n\log_2\frac{p}{2\sqrt{p-1}}-\log_2(2p),
\end{equation}
which is Eq.~\eqref{eq:qlower} of the main text. Sending the entire
string uses $n\log_2p+O(1)$ qubits, establishing the corresponding
linear upper bound. Public randomness cannot weaken the exact lower
bound: since the input and matching sets are finite, zero error for
every input permits fixing a public seed on which the protocol is
correct for all of them. The communication retrieval proof also applies to each fixed cyclic alphabet $\mathbb Z_q$ with $q>2$, using the same multiplicity bound; primality is needed for the orthogonality criterion in Appendix~\ref{app:phase}.\par

\smallskip\noindent\textit{(c) Encoding-independent lower bound for exact exclusion:}\enspace
We now prove the exclusion dimension lower bound, without assuming a phase encoding
or a specific measurement architecture. We give the operator-level details underlying Eqs.~\eqref{eq:em-weight}--\eqref{eq:em-average} of the main text.
The message dimension bounds the total weight at nonzero frequencies,
and exact exclusion forces each matching to carry at least $1/(p-1)$
of this weight on its edge frequencies.
Let $\rho_x$ be any
$d$-dimensional density operator assigned to input $x\in\Fp^n$.
Write
\begin{equation}
\bar\rho=\E_x\rho_x,\qquad
R_r=\E_x[\omega^{-r\cdot x}\rho_x],\qquad
s=\operatorname{rank}\bar\rho\le d.
\label{eq:supp-quantum-fourier}
\end{equation}
All $\rho_x$ have support in $\supp\bar\rho$, because their
uniform average is $\bar\rho$ and each is positive. Throughout,
$\bar\rho^{-1}$ denotes the inverse restricted to this support.
Define the nonnegative weights
\begin{equation}
w_r=\Tr(R_r^\dagger\bar\rho^{-1}R_r).
\label{eq:supp-quantum-weights}
\end{equation}
By matrix-valued Parseval,
\begin{align}
\sum_{r\in\Fp^n}w_r
&=\E_x\Tr(\rho_x\bar\rho^{-1}\rho_x)\\
&=\Tr(\bar\rho^{-1}\E_x\rho_x^2)
\le\Tr(\bar\rho^{-1}\bar\rho)=s,
\label{eq:supp-quantum-parseval}
\end{align}
where $\rho_x^2\le\rho_x$ for each density operator.
Since $R_0=\bar\rho$, we have $w_0=1$, whence
\begin{equation}
\sum_{r\ne0}w_r\le s-1.
\label{eq:supp-quantum-nonconstant}
\end{equation}

For a fixed perfect matching $M$, let $\{E_{M,e,b}\}_{e\in M,b\in\Fp}$
be Bob's POVM. Since every reported label must differ from the actual
edge difference, exactness says
\begin{equation}
\sum_{e\in M,b}\E_x\bigl[
\mathbf1_{\{\Delta_e(x)=b\}}
\Tr(E_{M,e,b}\rho_x)\bigr]=0.
\label{eq:supp-quantum-forbidden}
\end{equation}
Put $v_e=\mathbf e_j-\mathbf e_i$ for $e=(i,j)$ with $i<j$.
Using
$\mathbf1_{\{\Delta_e=b\}}=p^{-1}\sum_{a\in\Fp}
\omega^{a(\Delta_e-b)}$ and multiplying
Eq.~\eqref{eq:supp-quantum-forbidden} by $p$, the $a=0$
contribution is
$\sum_{e,b}\Tr(E_{M,e,b}\bar\rho)=1$.
The remaining Fourier terms cancel it; since $a\mapsto-a$ permutes
$\Fp^*$, the triangle inequality implies
\begin{equation}
1\le\sum_{e\in M,b}\sum_{a\ne0}
\bigl|\Tr(E_{M,e,b}R_{av_e})\bigr|.
\label{eq:supp-quantum-abs}
\end{equation}
For every positive operator $E$ and every matrix $R$ on
$\supp\bar\rho$, Hilbert--Schmidt Cauchy--Schwarz gives
\begin{equation}
\bigl|\Tr(ER)\bigr|^2\le
\Tr(E\bar\rho)\,
\Tr(ER^\dagger\bar\rho^{-1}R).
\label{eq:supp-quantum-cs}
\end{equation}
Indeed, the product of the Hilbert--Schmidt norms of
$\bar\rho^{1/2}E^{1/2}$ and
$\bar\rho^{-1/2}RE^{1/2}$ bounds the absolute trace
of their product.
Applying scalar Cauchy--Schwarz to
Eq.~\eqref{eq:supp-quantum-abs}, the first factor is
\begin{equation}
\sum_{e,b,a\ne0}\Tr(E_{M,e,b}\bar\rho)
=(p-1)\Tr\bar\rho=p-1.
\end{equation}
For the second factor use
$\sum_bE_{M,e,b}\le I$ for each fixed $e$. Hence
\begin{align}
1&\le(p-1)\sum_{e\in M}\sum_{a\ne0}
\Tr(R_{av_e}^\dagger\bar\rho^{-1}R_{av_e})\\
&=(p-1)\sum_{e\in M}\sum_{a\ne0}w_{av_e}.
\label{eq:supp-quantum-matching}
\end{align}
The argument allows the POVM and the choice of output edge to depend
arbitrarily on $M$.

Average Eq.~\eqref{eq:supp-quantum-matching} over a uniformly random
perfect matching. Every unordered edge belongs to such a matching
with probability $1/(n-1)$. Since distinct edges have disjoint
nonzero Fourier lines $\{av_e:a\in\Fp^*\}$,
Eq.~\eqref{eq:supp-quantum-nonconstant} yields
\begin{equation}
1\le\frac{p-1}{n-1}\sum_{i<j}\sum_{a\ne0}w_{a(\mathbf e_j-\mathbf e_i)}
\le\frac{(p-1)(s-1)}{n-1}.
\label{eq:supp-quantum-final}
\end{equation}
Therefore
\begin{equation}
d\ge s\ge 1+\frac{n-1}{p-1}.
\end{equation}
Together with the explicit $n$-dimensional phase protocol, this
proves $Q_0^\rightarrow(\mathrm{DHAM}_{n,p})=\log_2 n+O_p(1)$
for each fixed odd prime $p$. For public-coin protocols, exact
success for each of finitely many input--matching pairs holds for
almost every public seed; intersect these full-measure sets and
fix a seed. This completes the proof of the bound.
\hfill$\square$
\section{Proof of Corollary~\ref{cor:phase}: phase-code retrieval and orthogonality}\label{app:phase}

\noindent\textbf{Corollary~\ref{cor:phase} (Phase-code retrieval and orthogonality).}
For uniform $x$, any nonzero decoding effect $G$, fixed edge $e$,
and label $b$ satisfy $\Pr(\Delta_e=b\mid G)\le2/p$.
The bound is attainable. Two phase states are orthogonal if and only
if each field symbol occurs equally often in their componentwise
differences.

\smallskip\noindent\textit{Proof.}---\enspace
Uniformly averaging the phase states removes every off-diagonal term:
\begin{equation}
\bar\rho=\E_x\ket{\psi_x}\bra{\psi_x}=\frac In.
\end{equation}
For a fixed edge $e=(i,j)$, conditioning on $\Delta_e(x)=b$ retains
only the coherence between its two endpoints:
\begin{equation}
\bar\rho_{e,b}=\frac1n
\left(I+\omega^{-b}\ket i\bra j
+\omega^b\ket j\bra i\right).
\label{eq:supp-conditional-state}
\end{equation}
On $\operatorname{span}\{\ket i,\ket j\}$, this operator has
eigenvalues $0$ and $2/n$; all other eigenvalues are $1/n$.
Hence $\bar\rho_{e,b}\le2I/n$. If a positive decoding effect $G\ne0$
is observed, Bayes' rule and the uniform prior
$\Pr(\Delta_e=b)=1/p$ give
\begin{equation}
\Pr(\Delta_e=b\mid G)
=\frac{\Tr(G\bar\rho_{e,b})/p}{\Tr(G\bar\rho)}
\le\frac2p.
\label{eq:supp-confidence}
\end{equation}
This applies to every effect, not merely to a particular POVM. The
upper eigenvector is
$\ket{\phi_{e,b}}=(\ket i+\omega^b\ket j)/\sqrt2$.
For each perfect matching, the effects
$(2/p)\ket{\phi_{e,b}}\bra{\phi_{e,b}}$, indexed by its edges and
labels, sum to $I$ and saturate the bound on average. Thus $2/p$
is the optimal retrieval success \emph{for this encoding}; it is not
an encoding-independent success bound.

Finally, putting $z=y-x$ gives
\begin{equation}
\langle\psi_x|\psi_y\rangle
=\frac1n\sum_{a\in\Fp}N_a(z)\omega^a.
\end{equation}
The integer polynomial $\sum_{a=0}^{p-1}N_a(z)t^a$ has degree at most
$p-1$. Since $p$ is prime, the minimal polynomial of a primitive
$p$th root of unity is $1+t+\cdots+t^{p-1}$. Its value at $\omega$
vanishes only if all $N_a(z)$ are equal. Their sum is $n$, so each
must equal $n/p$. This also shows that orthogonal phase-state pairs
can occur only if $p$ divides $n$. \hfill$\square$
\section{Preparation-noncontextual posterior reduction}\label{app:ontic}

\begin{lemma}[Fourier form of an ontic posterior]\label{lem:posterior}
Under the operational equivalences in Eq.~\eqref{eq:equivs} of the
main text, the posterior of a uniformly sampled input, conditioned on
almost any ontic state, has the nonnegative pair-difference Fourier
form in Eq.~\eqref{eq:posterior} of the main text. Conversely, every
normalized nonnegative posterior with this Fourier support can be
implemented by a parity-oblivious classical encoding.
\end{lemma}

\smallskip\noindent\textit{Proof.}---\enspace
We first check that the quantum phase preparations satisfy the
assumed operational equivalences. Expanding the density operator
gives
\begin{equation}
\rho_x=\frac1n\sum_{i,j=1}^n
\omega^{x_i-x_j}\ket i\bra j.
\end{equation}
The diagonal term is $I/n$. Each off-diagonal term carries a
pair-difference Fourier character. For any forbidden direction $r$,
its average over the affine class $r\cdot x=k$ vanishes, since no
nonzero multiple of $r$ is a pair difference. Consequently,
$\overline\rho_{r,k}=I/n$ for every $k$, as asserted in
Eq.~\eqref{eq:equivs} of the main text.

Now consider the classical probability distribution of inputs
conditioned on an ontic state. A convenient normalization is
$f_\lambda(x)=p^n\Pr(x\mid\lambda)$, so
$f_\lambda\ge0$ and $\E_xf_\lambda(x)=1$. For $X$ uniform on $\Fp^n$ and any function $f$ on
$\Fp^n$, Fourier inversion gives, for nonzero $r$,
\begin{equation}
\E[f(X)\mid r\cdot X=k]
=\sum_{a\in\Fp}\widehat f(ar)\omega^{ak}.
\label{eq:supp-affine-fourier}
\end{equation}
If the conditional mean is independent of $k$, the Fourier
coefficients $\widehat f(ar)$ vanish for every $a\ne0$.
Applying this condition to every forbidden direction removes all
nonconstant Fourier characters except those associated with pair
differences. Grouping the surviving characters edge by edge gives
exactly Eq.~\eqref{eq:posterior} of the main text, with
$\sum_bh_{ij,\lambda}(b)=0$. Distinct edge lines have no nonzero
characters in common, so the decomposition is unambiguous.

The argument also holds without assuming a discrete or finite ontic
space. Write $\mu_x$ for the probability measure representing
preparation $x$, and let
\begin{equation}
\nu=p^{-n}\sum_{x\in\Fp^n}\mu_x.
\end{equation}
Every $\mu_x$ is absolutely continuous with respect to $\nu$.
The Radon--Nikodym theorem~\cite{Folland1999} therefore gives a derivative
$w_x=d\mu_x/d\nu$. Bayes' rule then gives, for $\nu$-almost all
$\lambda$,
\begin{equation}
\Pr(x\mid\lambda)=p^{-n}w_x(\lambda),\qquad
f_\lambda(x)=w_x(\lambda).
\end{equation}
Preparation noncontextuality equates the ontic mixtures associated
with every pair of operationally equivalent affine classes.
Taking their Radon--Nikodym derivatives yields
\begin{equation}
\sum_{x:r\cdot x=k}\Pr(x\mid\lambda)
=\sum_{x:r\cdot x=k'}\Pr(x\mid\lambda).
\end{equation}
Only finitely many inputs and equivalences are involved, so all
these equalities hold outside a single $\nu$-null set.
Equation~\eqref{eq:supp-affine-fourier} then applies to
$f_\lambda$ for almost every $\lambda$. This establishes the
posterior reduction for arbitrary measurable ontic spaces.

To prove the converse, start from any nonnegative normalized function
$f$ with the permitted Fourier support. Introduce classical messages
$t\in\Fp^n$ and define
\begin{equation}
\Pr(t\mid x)=p^{-n}f(x-t).
\label{eq:twirl}
\end{equation}
The probabilities sum to one because translating $t$ runs over all
arguments of $f$. For uniform inputs, $t$ is itself uniform and the
posterior given $t$ is $p^{-n}f(x-t)$. Translation only changes
Fourier phases and therefore preserves all forbidden operational
equivalences. Any exclusion label $b$ for the shifted string $x-t$
is converted to a label for $x$ by adding $t_j-t_i$ on the chosen
edge $e=(i,j)$. Thus the classical protocol has exactly the success
score associated with $f$. \hfill$\square$
\section{Triangle obstruction and its operational meaning}\label{app:triangle}

\noindent\textbf{Theorem~\ref{thm:triangle} (Triangle obstruction).}
For any prime $p$, the edges supporting conclusive
exclusion under a posterior of the form
\eqref{eq:posterior} lie within a single triangle.
Consequently, for every even $n\ge6$, no
preparation-noncontextual strategy can win
perfectly over all matchings, whereas the quantum
protocol does. For $n=4$, a perfect parity-oblivious
classical strategy exists.

\smallskip\noindent\textit{Proof.}---\enspace
Let $f$ be any allowed normalized posterior, and set
\begin{equation}
r_e(b)=\Pr_f(\Delta_e=b)
=\frac{1+h_e(b)}p,
\end{equation}
where $h_e$ denotes the corresponding edge term in the Fourier
representation of Eq.~\eqref{eq:posterior} of the main text.
The formula follows by averaging over all coordinates except the
difference on $e$: every other edge term has mean zero. If $b$ is
excluded on $e$, then $r_e(b)=0$ and $h_e(b)=-1$. Since $f\ge0$, a
zero marginal also forces $f$ to vanish at every point of the affine
hyperplane $\Delta_e=b$.

\smallskip\noindent\textit{Disjoint excluded edges are impossible:}\enspace
Suppose two vertex-disjoint edges $e,e'$ exclude values $b,b'$.
Conditioning simultaneously on their differences, the two
corresponding edge terms are fixed at $-1$. All other edge terms
average to zero, including terms joining the two edges, because at
least one still-uniform endpoint remains. Therefore
\begin{equation}
\E[f\mid\Delta_e=b,\Delta_{e'}=b']
=1+h_e(b)+h_{e'}(b')=-1,
\end{equation}
in contradiction with $f\ge0$. Thus any two excluded edges share a
vertex.

\smallskip\noindent\textit{A three-star of excluded edges is also impossible:}\enspace
Suppose three edges $(0,1)$, $(0,2)$, $(0,3)$ exclude differences
$b_1,b_2,b_3$. Average $f$ over the other coordinates and use the
relative variables $y_j=x_j-x_0$. The resulting normalized,
nonnegative function $F(y_1,y_2,y_3)$ contains only constant,
one-variable, and two-variable terms: every surviving Fourier
character comes from a pair difference. Since the three values are
excluded, $F$ vanishes whenever any $y_j=b_j$.

Define the finite-difference operator
\begin{equation}
(D_jF)(y)=F(y)-F(y_1,\ldots,y_{j-1},b_j,y_{j+1},\ldots).
\end{equation}
The vanishing condition gives $D_jF=F$ for each $j$, hence
$D_1D_2D_3F=F$. But every term of $F$ depends on at most two of
the variables, so its third mixed difference is zero. This forces
$F=0$, contrary to its normalization. A three-star is therefore
excluded.

A graph in which every pair of edges intersects is contained in a
star or a triangle. Since a three-star is impossible, the entire
excluded-edge graph lies in a triangle. This proves the structural
claim.

For even $n\ge6$, every triangle can be avoided by some perfect
matching: match each of its three vertices to a distinct vertex
outside the triangle, then match the remaining vertices among
themselves. Therefore no posterior can exclude a value on an edge
of \emph{every} perfect matching. By contrast, every perfect
matching on four vertices intersects any fixed triangle.

For completeness, we explicitly construct an allowed posterior that
excludes a value on all three edges of a triangle. Choose vertices
$1,2,3$, a nonzero $t\in\Fp$, and put
$u=x_2-x_1$, $v=x_3-x_2$. Define
\begin{equation}
A(s)=\ind_{s=t}-\ind_{s=0},\qquad
f_\triangle(x)=1+A(u)+A(v)-A(u+v).
\label{eq:triangleposterior}
\end{equation}
This function has mean one and contains only the three allowed pair
differences. If $u=0$ or $v=0$, its value is zero; if $u+v=t$,
its value is also zero. Otherwise $u,v\ne0$ and $u+v\ne t$, so all
remaining terms are nonnegative and $f_\triangle\ge1$.
In particular $f_\triangle\ge0$ everywhere. It excludes a value
on each of the edges $\{1,2\}$, $\{2,3\}$, and $\{1,3\}$.
By the translation construction of Appendix~\ref{app:ontic}, Eq.~\eqref{eq:twirl}, it yields a perfect parity-oblivious
classical protocol for any matching family intersected by that
triangle. This includes all perfect matchings when $n=4$.

\smallskip\noindent\textit{Consequence for matching families:}\enspace
Suppose a preparation-noncontextual strategy succeeds with
probability one on every matching in a finite family $\mathcal S$.
For almost every ontic state, its conditional failure probability
must then vanish on every setting, since a nonnegative function
with zero average is zero almost everywhere. The triangle
obstruction therefore supplies a single triangle meeting all
matchings in $\mathcal S$. Conversely,
Eq.~\eqref{eq:triangleposterior} constructs a perfect strategy
whenever such a triangle exists. This proves the matching-family
criterion used in Theorem~\ref{thm:settings}. No measurement noncontextuality or
outcome determinism has been assumed. \hfill$\square$
\section{Matching-family criterion and the optimal three-setting witness}\label{app:three}
\smallskip\noindent\textbf{Finite-dimensional separation for the three settings.}
For $n=6$, write the three matchings as
$M_0=\{14,25,36\}$, $M_1=\{15,26,34\}$, and
$M_2=\{16,24,35\}$. Consider the $p^2$ inputs
$x^{(u,v)}=(0,u,2u,v,u+v,2u+v)$, with
$(u,v)\in\mathbb F_p^2$. For two distinct inputs,
let $a=u-u'$ and $b=v-v'$. If $b\ne0$, the matched
differences between the two inputs on $M_0$ are
$(b,b,b)$, all nonzero. If $b=0$, then $a\ne0$,
and the matched differences on $M_1$ are
$(a,a,-2a)$, again all nonzero for odd $p$.
Thus, for each pair of distinct inputs, some allowed
matching has disjoint sets of correct retrieval outcomes.
Any zero-error retrieval POVM on that matching perfectly
distinguishes the corresponding message states, so their
supports are orthogonal. All $p^2$ message states must
therefore have mutually orthogonal supports, proving
$d_{\mathrm{DHM}}\ge p^2$ for arbitrary mixed-state
encodings and POVMs. The phase protocol in the main
text achieves exact DHAM exclusion in dimension $n=6$.
These bounds do not assert that either restricted-family
dimension is optimal.\par
\vspace{0.2cm}

\noindent\textbf{Theorem~\ref{thm:settings} (Matching-family characterization and optimal three-setting witness).}
A family $\mathcal S\subseteq\Mn$ admits perfect
parity-oblivious classical, equivalently
preparation-noncontextual, exclusion if and only
if some triangle contains an edge of every
$M\in\mathcal S$. For every even $n\ge6$,
three matching settings are necessary and
sufficient for an AVN contradiction.
Specifically, for $n=2m$, consider
\begin{equation*}
M_s=\{\{a_r,b_{r+s}\}:r\in\mathbb Z_m\},
\quad s=0,1,2,
\end{equation*}
where $\{a_r\}$ and $\{b_r\}$ are disjoint
vertex sets. Their uniformly averaged optimal
noncontextual exclusion score is
\begin{equation*}
P_{\NC}^{(3)}(n,p)
=1-\min\!\left\{\frac1{3p},\frac1{p^2}\right\}.
\end{equation*}
This gives $5/6$ for $p=2$ and $1-1/p^2$
for odd primes, against unit quantum success.

\smallskip\noindent\textit{Proof.}---\enspace
We establish the characterization and minimality before optimizing
the finite three-setting score.

\smallskip\noindent\textit{(a) Characterization and minimal number of settings:}\enspace
The necessity of a triangle meeting each matching follows from
Theorem~\ref{thm:triangle} (Appendix~\ref{app:triangle}). For sufficiency, the triangle posterior in
Eq.~\eqref{eq:triangleposterior} excludes a prescribed difference
on each of its edges; Lemma~\ref{lem:posterior} converts it into a parity-oblivious
classical protocol. Thus the triangle criterion is both necessary
and sufficient.

To see why two settings never suffice for an AVN contradiction,
choose an edge $\{u,v\}$ of the first matching. If $u$ is paired
with $w\ne v$ in the second, the triangle on $\{u,v,w\}$ meets both
matchings. If the two matchings share $\{u,v\}$, any triangle
containing that edge suffices. Hence every pair of perfect
matchings has a triangle transversal.

Now take the three cyclic matchings $M_0,M_1,M_2$ in
Eq.~\eqref{eq:settings} of the main text. Since $m=n/2\ge3$,
they are pairwise edge-disjoint. Their union is bipartite, with
parts $\{a_r\}$ and $\{b_r\}$; it contains no triangle. A triangle
meeting all three matchings would require three distinct edges,
one from each setting, and hence would form a triangle in the
union, which is impossible. Therefore these three settings give
an AVN contradiction, and three are minimal.

\smallskip\noindent\textit{(b) Lower bound on the average failure probability:}\enspace
Fix an allowed normalized posterior $f$ and define
\begin{equation}
m_e=\min_{b\in\Fp}\Pr_f(\Delta_e=b).
\end{equation}
For a given matching, Bob minimizes his failure probability by
selecting an edge with smallest $m_e$ and reporting a label that
attains this minimum. Choose one such edge $e_s\in M_s$ for
$s=0,1,2$. The optimal averaged failure for this posterior is
\begin{equation}
F(f)=\frac{m_{e_0}+m_{e_1}+m_{e_2}}3.
\label{eq:supp-three-failure}
\end{equation}

If two selected edges $e,e'$ are disjoint, let $b,b'$ minimize their
respective marginals. Just as in the proof of Theorem~\ref{thm:triangle},
nonnegativity gives
\begin{equation}
0\le\E[f\mid\Delta_e=b,\Delta_{e'}=b']
=p(m_e+m_{e'})-1.
\end{equation}
Since the third marginal is nonnegative,
Eq.~\eqref{eq:supp-three-failure} implies
$F(f)\ge1/(3p)$.

Otherwise all three chosen edges intersect pairwise. Their union
cannot be a triangle because it is contained in the bipartite
union of the three cyclic matchings. It must therefore be a
three-star: three edges sharing one vertex.

For $p\ge3$, label the central vertex $0$ and its leaves $1,2,3$.
Translation of each $y_j=x_j-x_0$ makes the minimizing values zero.
Average over other coordinates, then symmetrize over permutations
of the leaves and common multiplication of all $y_j$ by a nonzero
field element. These operations preserve nonnegativity, permitted
Fourier support, and the sum of the three minimizing marginals.
Each surviving edge term depends on $y_j$ or on $y_i-y_j$.
Common scaling averages its values over all nonzero arguments, leaving
only the distinction between zero and nonzero. Permuting the leaves
then gives one coefficient for the three central edges and one for
the three pairs of leaves.
Every symmetrized posterior has the form
\begin{equation}
f_{\mathrm{sym}}(y)=1+A(pN_0-3)+B(pN_{=}-3),
\label{eq:supp-symstar}
\end{equation}
where
\begin{equation}
N_0=\sum_{j=1}^3\ind_{y_j=0},\qquad
N_{=}=\sum_{1\le i<j\le3}\ind_{y_i=y_j}.
\end{equation}
The first term counts zero leaf differences; the second counts
equal pairs of leaf differences. By symmetry, each star edge has
the same probability of difference zero,
\begin{equation}
m=\Pr_{f_{\mathrm{sym}}}(y_j=0)
=\frac{1+(p-1)A}{p}.
\end{equation}
Positivity at $(y_1,y_2,y_3)=(0,0,t)$, $t\ne0$, and at
$(0,t,u)$ for distinct nonzero $t,u$ yields
\begin{align}
1+(2p-3)A+(p-3)B&\ge0,\\
1+(p-3)A-3B&\ge0.
\label{eq:supp-star-constraints}
\end{align}
Multiply the first inequality by $3$ and the second by $p-3$.
The $B$ terms cancel, leaving $p+p^2A\ge0$.
Hence $A\ge-1/p$ and $m\ge1/p^2$.
Because the symmetrization preserves the sum of the three marginals,
Eq.~\eqref{eq:supp-three-failure} gives $F(f)\ge1/p^2$.

For $p=2$, the three-star case gives an average failure probability of at least $1/6$. Put
$z_j=(-1)^{y_j}$ and expand the averaged posterior as
\begin{equation}
F(z)=1+\sum_{j=1}^3a_jz_j
+\sum_{1\le i<j\le3}b_{ij}z_iz_j.
\end{equation}
After shifting the minimizing differences to zero,
$m_j=(1+a_j)/2$. Sum the four positivity inequalities at
$(1,1,1)$ and at the three assignments with exactly one minus
sign. All quadratic terms cancel, giving
$4+2\sum_ja_j\ge0$. Therefore
$\sum_jm_j\ge1/2$, and the average failure is at least
$1/6=1/(3p)$.

Combining the disjoint-edge and three-star cases proves
\begin{equation}
F(f)\ge\min\left\{\frac1{3p},\frac1{p^2}\right\},
\end{equation}
which gives the upper bound on success in
Eq.~\eqref{eq:3bound} of the main text.

\smallskip\noindent\textit{(c) Stronger characterized-phase-ensemble bound:}\enspace
For the characterized phase ensemble, every nonconstant Fourier
coefficient of an ontic posterior lies on a first-harmonic
pair-difference frequency $\pm(\mathbf e_j-\mathbf e_i)$.
Indeed, every absent character defines a signed linear relation
among the preparations. For each absent nonconstant
frequency, the real and imaginary parts of the Fourier relation
give real zero-sum relations among the preparations. Normalizing
the positive and negative parts of each nontrivial relation gives
two operationally equivalent mixtures, so preparation
noncontextuality imposes the same relation on the ontic
distributions. The posterior reduction in Appendix~\ref{app:ontic}
then transfers this restricted Fourier support to each posterior.

Choose in each of the three cyclic settings an edge and exclusion
label minimizing failure, as in part (b). If two chosen edges
are disjoint, Eq.~\eqref{eq:supp-three-failure} and the disjoint-edge
argument give average failure at least $1/(3p)$. Otherwise the
three edges form a star. Orient the leaf differences away from
the common centre and translate their minimizing labels to zero.
Average over all coordinates outside the four star
vertices. Average over permutations of the leaves and the common reflection
$y\mapsto-y$. These transformations preserve the mean failure
and first-harmonic Fourier support. The resulting normalized,
nonnegative posterior has the form
\begin{equation}
F(y)=1+A\sum_{j=1}^{3}c(y_j)
+B\sum_{1\le i<j\le3}c(y_i-y_j),
\quad c(t)=\cos(2\pi t/p).
\label{eq:supp-phase-star}
\end{equation}
For each chosen star edge the failure probability is
$m=(1+A)/p$. At $y=(0,0,t)$, positivity implies
\begin{equation}
1+A[2+c(t)]+B[1+2c(t)]\ge0.
\end{equation}
For odd $p$, the values of $c(t)$ include $1$ and
$-\cos(\pi/p)\le-1/2$. Thus a convex combination of
these inequalities can be chosen with mean $c(t)=-1/2$.
The $B$ coefficient cancels, leaving $1+3A/2\ge0$.
Hence $m\ge1/(3p)$, and the mean failure probability
over the three cyclic settings is therefore at least $1/(3p)$. Therefore
\begin{equation}
P_{\NC,\mathrm{phase}}^{(3)}\le1-\frac1{3p}.
\label{eq:supp-phase-bound}
\end{equation}
For $p\ge5$, this is strictly smaller than
$1-1/p^2$, proving nonsaturation of the stipulated
bound under all characterized-phase-ensemble equivalences.
This is Eq.~\eqref{eq:phase-nc} of the main text; no claim of tightness is made.

\smallskip\noindent\textit{(d) Saturating strategies for the stipulated equivalences:}\enspace
For odd $p$, choose one star edge from each of the three settings
and write
\begin{equation}
f_\star(x)=1-\sum_{j=1}^3\ind_{x_j=x_0}
+\sum_{1\le i<j\le3}\ind_{x_i=x_j}.
\label{eq:supp-star-posterior}
\end{equation}
The posterior has mean one and contains only allowed pair
differences. If $k$ of the leaves equal the center, at least
$\binom k2$ leaf pairs coincide. Thus
$f_\star\ge1-k+\binom k2\ge0$ for $k=0,1,2,3$.
On each selected star edge, the probability of the excluded label
zero is exactly $1/p^2$. Bob uses that edge and label in each
setting, attaining success $1-1/p^2$.

For $p=2$, choose a triangle meeting the first two cyclic
matchings. Its third edge lies within one bipartition class and
therefore in none of the three matchings. The triangle posterior
in Eq.~\eqref{eq:triangleposterior} succeeds perfectly on the
first two settings and gives uniform differences on every edge
of the third, for which exclusion succeeds with probability
$1/2$. The average success is $(1+1+1/2)/3=5/6$.
By the translation averaging of Lemma~\ref{lem:posterior}, both posteriors define
operational parity-oblivious classical protocols. The lower bounds
are therefore tight, completing Theorem~\ref{thm:settings}. \hfill$\square$
\section{Noncontextual scaling over the full matching family}\label{app:full}

\noindent\textbf{Theorem~\ref{thm:full} (Full-family noncontextual scaling).}
For every fixed prime $p$,
\begin{align*}
P_{\NC}^{\excl}(n,p)&=1-\frac1p+\Theta_p(n^{-1/2}),
\\
P_{\NC}^{\ret}(n,p)&=\frac1p+\Theta_p(n^{-1/2}).
\end{align*}
For every even $n\ge6$,
$P_{\NC}^{\excl}(n,p)\le
1-\min\{1/(3p),1/p^2\}$.

\smallskip\noindent\textit{Proof.}---\enspace
We show matching upper and lower orders for the advantage, and then
establish the finite-size bound.

\smallskip\noindent\textit{(a) Universal upper bound:}\enspace
Let $f$ be an allowed nonnegative posterior with $\E f=1$, and
put $r_e(b)=\Pr_f(\Delta_e=b)$. Define
\begin{equation}
\delta_e=\frac1p-\min_br_e(b),\qquad
\gamma_e=\max_br_e(b)-\frac1p.
\end{equation}
When Bob chooses the best edge and label in a given matching,
his conditional scores are
\begin{equation}
S_{\excl}(f,M)=1-\frac1p+\max_{e\in M}\delta_e,
\qquad
S_{\ret}(f,M)=\frac1p+\max_{e\in M}\gamma_e.
\label{eq:supp-full-scores}
\end{equation}

The Fourier support in Eq.~\eqref{eq:posterior} of the main text
has Hamming degree at most two. Let $T_\rho$ multiply Fourier
characters of degree $k$ by $\rho^k$. Finite-symbol
hypercontractivity gives a constant $0<\rho_p<1$, depending only
on $p$, such that
\begin{equation}
\|T_{\rho_p}g\|_2\le\|g\|_{4/3}
\end{equation}
for every function on $\Fp^n$~\cite{wolff}. Since $f$ has
Fourier degree at most two, Parseval's identity and $\rho_p<1$
yield
\begin{equation}
\|f\|_2\le\rho_p^{-2}\|T_{\rho_p}f\|_2
\le\rho_p^{-2}\|f\|_{4/3}.
\end{equation}
Interpolation between $L_1$ and $L_2$, using $f\ge0$ and
$\|f\|_1=1$, gives $\|f\|_{4/3}\le\|f\|_2^{1/2}$.
Thus $\|f\|_2\le\rho_p^{-4}$. The Fourier subspaces associated
with distinct edges are orthogonal, so
\begin{equation}
\sum_e\|h_e\|_2^2=\|f-1\|_2^2
\le\rho_p^{-8}-1.
\label{eq:supp-full-energy}
\end{equation}
Here $\|h_e\|_2$ is normalized with respect to the uniform
difference on $\Fp$.

Because $r_e(b)=[1+h_e(b)]/p$ and $h_e$ has zero mean,
Cauchy--Schwarz applied to the remaining $p-1$ values gives
\begin{equation}
\delta_e^2,\ \gamma_e^2
\le\frac{p-1}{p^2}\|h_e\|_2^2.
\end{equation}
Together with Eq.~\eqref{eq:supp-full-energy}, this implies
\begin{equation}
\sum_e\delta_e^2,\quad\sum_e\gamma_e^2
\le B_p:=\frac{p-1}{p^2}(\rho_p^{-8}-1).
\label{eq:supp-full-bonus}
\end{equation}
Each edge occurs in a uniformly random matching with probability
$1/(n-1)$. Bounding a maximum by the $\ell_2$ norm and using
Cauchy--Schwarz once more gives
\begin{equation}
\E_M\max_{e\in M}\delta_e,\quad
\E_M\max_{e\in M}\gamma_e
\le\sqrt{\frac{B_p}{n-1}}.
\end{equation}
The same bound survives averaging over ontic states, since $B_p$
is independent of the posterior. This proves the
$O_p(n^{-1/2})$ contributions in Eq.~\eqref{eq:fullbound} of the main text.

\smallskip\noindent\textit{(b) Construction attaining the scaling:}\enspace
Fix a subset $K\subseteq[n]$ of $k$ vertices and set
\begin{equation}
f_K(x)=1+\lambda\sum_{\{i,j\}\subset K}
\bigl(p\ind_{x_i=x_j}-1\bigr),\qquad
\lambda=\frac2{(p-1)k}.
\label{eq:supp-clique}
\end{equation}
Every term has zero mean and uses an allowed pair-difference
character, so $f_K$ is normalized and has the required Fourier
support. To verify positivity, let $N_a$ count the symbols equal
to $a$ among the coordinates in $K$. Counting equal-symbol pairs
gives
\begin{align}
&\sum_{\{i,j\}\subset K}
\bigl(p\ind_{x_i=x_j}-1\bigr)\\
&\qquad=\frac12\left(p\sum_aN_a^2-k^2-(p-1)k\right)
\ge-\frac{(p-1)k}{2},
\end{align}
where the last inequality uses
$\sum_aN_a^2\ge k^2/p$. With the chosen $\lambda$ this shows
$f_K\ge0$.

For an edge wholly inside $K$, its difference distribution is
\begin{equation}
r_e(0)=\frac{1+(p-1)\lambda}{p},\qquad
r_e(b\ne0)=\frac{1-\lambda}{p}.
\end{equation}
Consequently, such an edge gives exclusion bonus $\lambda/p$
and retrieval bonus $(p-1)\lambda/p$. A uniformly random
perfect matching contains an edge inside $K$ with probability
$1-q(n,k)$, where $q(n,k)$ is the no-full-edge probability
given in Eq.~\eqref{eq:sample-noedge} of Appendix~\ref{app:classical}, part (c). Choosing
$k=\lfloor\sqrt n\rfloor$ gives
$1-q(n,k)\to1-e^{-1/2}>0$ and $\lambda=\Theta_p(n^{-1/2})$.
This realizes an $\Omega_p(n^{-1/2})$ bonus for both tasks.
Together with the upper bound, it proves the tight asymptotic
claims in Eq.~\eqref{eq:fullbound} of the
main text.

\smallskip\noindent\textit{(c) Uniform finite-size exclusion bound.}\enspace
Apply the tight three-setting bound of Theorem~\ref{thm:settings} to any allowed
posterior and average over a uniformly random relabeling of the
$n$ vertices. Every relabeled setting is then individually
uniform over all perfect matchings. By linearity of expectation,
the average full-family failure is at least
$\min\{1/(3p),1/p^2\}$. Hence, for every even $n\ge6$,
\begin{equation}
P_{\NC}^{\excl}(n,p)
\le1-\min\left\{\frac1{3p},\frac1{p^2}\right\}.
\end{equation}
This is the finite-$n$ inequality in Theorem~\ref{thm:full}. \hfill$\square$
\section{Exclusion effects and projective-simulable measurements}\label{app:measurement}

\noindent\textbf{Proposition~\ref{prop:measurement} (Conclusive readout and measurement simulability).}
For the phase preparations, any nonzero effect $F_{e,b}$ that perfectly excludes $b$ whenever $\Delta_e=b$ must satisfy
\begin{equation*}
\supp F_{e,b}\subseteq
\operatorname{span}\{\ket{\chi_{e,b}}\}.
\end{equation*}
For odd $p$, perfect edge-resolved exclusion requires at least three nonzero effects; for $p=3$, the label-aggregated effects uniquely form the antitrine POVM. For odd $p$, with a characterized pure phase source, uniform inputs, and characterized edge projection, every projective-simulable qubit measurement satisfies
\begin{equation*}
S_{\excl}\le S_{\PS}(p)
=1-\frac{1-\cos(\pi/2p)}p.
\end{equation*}

\smallskip\noindent\textit{Proof.}---\enspace
We separate the exact exclusion constraint from the
projective-simulability optimization, since they use different
properties of the readout.

\smallskip\noindent\textit{(a) Rigidity of perfect exclusion:}\enspace
Fix an edge $e=(i,j)$ and a label $b$. The conditional state
$\bar\rho_{e,b}$ is given in Eq.~\eqref{eq:supp-conditional-state}
in Appendix~\ref{app:phase}. It has a one-dimensional kernel,
spanned by
\begin{equation}
\ket{\chi_{e,b}}=\frac{\ket i-\omega^b\ket j}{\sqrt2}.
\end{equation}
If a positive effect $F_{e,b}$ excludes $b$ with certainty, then
$\Tr(F_{e,b}\bar\rho_{e,b})=0$. Positivity of both operators
implies that $F_{e,b}$ is supported in the kernel of the
conditional state. This proves the support condition in
Eq.~\eqref{eq:rigidity} of the main text, and also shows that an
effect excluding two different labels must be zero because the
corresponding kernel rays are distinct.

For a qubit, two nonzero rank-one positive effects can sum to the
identity only if their rays are orthogonal. Here
\begin{equation}
\langle\chi_{e,a}|\chi_{e,b}\rangle
=\frac{1+\omega^{b-a}}2,
\end{equation}
which vanishes only if $\omega^{b-a}=-1$. No such pair exists
for odd $p$, so at least three nonzero effects are required.
Three suffice: take $a=(p-1)/2$, $b=(p+1)/2$ and
$c=\cos(\pi/p)$. The effects
\begin{align}
F_{e,0}&=\frac{2c}{1+c}
\ket{\chi_{e,0}}\bra{\chi_{e,0}},\\
F_{e,a}&=\frac1{1+c}
\ket{\chi_{e,a}}\bra{\chi_{e,a}},\\
F_{e,b}&=\frac1{1+c}
\ket{\chi_{e,b}}\bra{\chi_{e,b}}
\end{align}
sum to the edge identity. Geometrically, three exclusion rays
resolve a qubit identity when the origin lies in the convex hull
of their Bloch directions. For $p=3$, all three labels must be
used and completeness fixes each coefficient to $2/3$.
The label-aggregated POVM is therefore the antitrine measurement;
splitting an effect into proportional effects bearing the same
label does not change this conclusion.

\smallskip\noindent\textit{(b) Optimal projective-simulable bound:}\enspace
Assume the phase source and the edge projection are
characterized. Conditioned on the edge, the possible qubit states
are $\rho_b=\ket{\phi_b}\bra{\phi_b}$, where
$\ket{\phi_b}=(\ket i+\omega^b\ket j)/\sqrt2$ and $b$ is uniform.
For a POVM $\{M_b\}_{b\in\Fp}$ used to output the excluded label,
its average failure probability is
\begin{equation}
F=\frac1p\sum_{b\in\Fp}\Tr(\rho_bM_b).
\label{eq:supp-measurement-failure}
\end{equation}

Projective-simulable qubit POVMs are convex mixtures of projective
measurements followed by classical post-processing. Since
Eq.~\eqref{eq:supp-measurement-failure} is linear in the POVM,
there is an optimal strategy consisting of a single binary
projective measurement with a deterministic output label assigned
to each outcome. If the two labels coincide, its failure is
$1/p$. Otherwise, call them $a\ne b$ and write the measurement
as $\{P,I-P\}$:
\begin{align}
F_{a,b}(P)
&=\frac1p\left[\Tr(\rho_aP)
+\Tr\bigl(\rho_b(I-P)\bigr)\right]\\
&=\frac1p\left[1+\Tr P(\rho_a-\rho_b)\right].
\end{align}
The Hermitian operator $\rho_a-\rho_b$ is traceless and has
eigenvalues $\pm\|\rho_a-\rho_b\|_1/2$. Choosing $P$ to
project onto its negative eigenspace minimizes the failure:
\begin{equation}
\min_PF_{a,b}(P)=\frac1p
\left(1-\frac12\|\rho_a-\rho_b\|_1\right).
\end{equation}
For these pure states,
\begin{equation}
\frac12\|\rho_a-\rho_b\|_1
=\sqrt{1-|\langle\phi_a|\phi_b\rangle|^2}
=\left|\sin\frac{\pi(a-b)}p\right|.
\end{equation}
For odd $p$, its largest value is
$\cos(\pi/2p)$, achieved by the two labels separated by
$(p\pm1)/2$ modulo $p$. Hence
\begin{equation}
S_{\excl}\le1-\frac{1-\cos(\pi/2p)}p=S_{\PS}(p).
\end{equation}
The optimizing binary projective measurement attains this bound,
so it is tight under the characterized-source assumptions.
At $p=3$, it equals $(4+\sqrt3)/6$.

\smallskip\noindent\textit{(c) Depolarization and distinct thresholds:}\enspace
Consider the depolarized preparation
$\rho_x(v)=v\ket{\psi_x}\bra{\psi_x}+(1-v)I/n$ used in the main
text. The ideal exclusion POVM never outputs the correct label on
the pure phase state. For the maximally mixed part it does so
with probability $1/p$. Its total success is therefore
\begin{equation}
S_Q(v)=1-\frac{1-v}{p},
\end{equation}
as in Eq.~\eqref{eq:noise} of the main text. For $p=3$,
comparison with the preparation-noncontextual bound $8/9$
yields $v>2/3$; comparison with the \emph{pure-phase}
projective-simulable score $(4+\sqrt3)/6$ yields
$v>\sqrt3/2$. To optimize for the calibrated noisy source, condition on the edge. Its state is $\rho_b(v)=v\rho_b+(1-v)I_2/2$, because the edge projection has probability $2/n$ on both components. Every pairwise trace distance in the preceding optimization is multiplied by $v$. For odd $p$ we therefore obtain the tight bound
\[
S_{\PS}(p,v)=1-\frac{1-v\cos(\pi/2p)}p .
\]
This reproduces Eq.~\eqref{eq:em-noisyps} of the main text.
The ideal ternary measurement exceeds this benchmark for every positive visibility. At $v=0$ all readouts have the random-guessing score. The comparison assumes a characterized source and edge projection. \hfill$\square$

\subsection{Noise robustness and quantitative resource-weight bounds}

\begin{figure}[htb]
\centering
\includegraphics[width=0.52\textwidth]{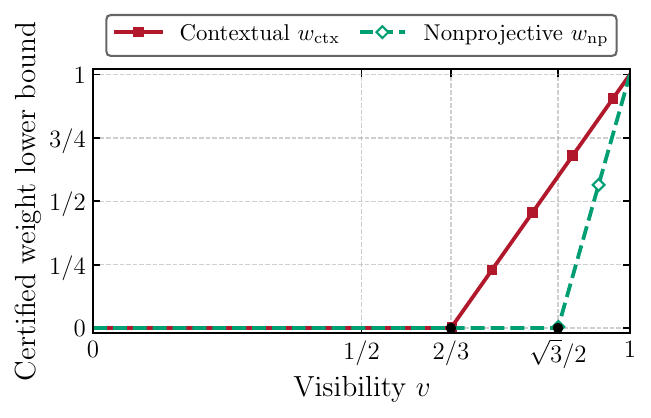}
\caption{\footnotesize
Certified lower bounds
$w_{\mathrm{ctx}}\ge\max\{0,3v-2\}$ and
$w_{\mathrm{np}}\ge\max\{0,(2v-\sqrt3)/(2-\sqrt3)\}$
for ternary depolarized phase preparations.
The measurement bound requires a characterized source and
edge-resolved readout.}
\label{fig:resource_weights}
\end{figure}

The observed score also bounds the minimum weight of a resourceful
component in a convex decomposition. Suppose a preparation or
measurement admits a decomposition into a free part of weight
$1-w$ and an unrestricted part of weight $w$. Since the score is
linear, the free part scores at most $S_{\mathrm{free}}$, and the
unrestricted score cannot exceed one,
\begin{equation}
S\le(1-w)S_{\mathrm{free}}+w.
\end{equation}
Solving this inequality for $w$ gives
\begin{equation}
w\ge\max\left\{0,
\frac{S-S_{\mathrm{free}}}{1-S_{\mathrm{free}}}\right\}.
\label{eq:supp-weight}
\end{equation}
For contextual weight, take
$S_{\mathrm{free}}=P_{\NC}^{(3)}=8/9$; for nonprojective
measurement weight, use
$S_{\mathrm{free}}=S_{\PS}(3)=(4+\sqrt3)/6$.
In the ternary depolarizing model $S_Q(v)=(2+v)/3$, giving
\begin{equation}
\begin{split}
w_{\mathrm{ctx}}&\ge\max\{0,3v-2\},\\
w_{\mathrm{np}}&\ge\max\left\{0,
\frac{2v-\sqrt3}{2-\sqrt3}\right\}.
\end{split}
\end{equation}
The second bound uses the conservative pure-source reference plotted in Fig.~\ref{fig:resource_weights}. With the calibrated ensemble fixed and $v>0$, the unrestricted measurement ceiling is $S_Q(v)$: every measurement scores $2/3$ on the white component, and the antitrine is perfect on the pure component. The stronger measurement-weight bound is therefore $\max\{0,[S-S_{\PS}(3,v)]/[S_Q(v)-S_{\PS}(3,v)]\}$.
These are bounds in distinct convex resource models: the first
assumes the stated preparation operational equivalences, while the
second additionally assumes a characterized phase source and an
edge-resolved qubit measurement. Their weights should not be
identified with one another.

\section{Finite ternary implementation and operational requirements}\label{app:implementation}
The smallest odd-prime AVN construction has $(n,p)=(6,3)$.
The ideal preparations
\begin{equation}
\ket{\psi_x}=\frac1{\sqrt6}\sum_{j=1}^{6}
\exp(2\pi i x_j/3)\ket j,
\qquad x\in\mathbb F_3^6,
\label{eq:s9-source}
\end{equation}
can in principle be encoded in six path or time-bin modes.
Although there are $3^6=729$ input labels, $x$ and
$x+c\mathbf1$ give the same density operator for any
$c\in\mathbb F_3$. There are therefore $3^5=243$ distinct
pure-state rays. A practical preparation procedure must preserve
these global-shift identifications operationally, rather than
assuming that an imperfect collection of phase modulators does so.

For $n=6$ Bob uses the three bipartite matching settings of
Fig.~\ref{fig:settings}. Each setting partitions the six modes into
three orthogonal two-mode subspaces. A which-pair projection followed
by the three-outcome antitrine POVM within the selected subspace
produces nine outcomes $(e,b)$ per matching. The complete
measurement has the effects of Eq.~\eqref{eq:povm}; the ideal
probability of each edge is $2/n=1/3$, independent of $x$.
Programmable six-mode interferometers are an experimentally relevant
architecture~\cite{Giordani2023}, while three-outcome qubit POVMs have
been realized with photonic quantum walks~\cite{Zhao2015}.
These works supply potential building blocks but do not test the
present DHAM witness.

The required preparation equivalences are the substantial overhead.
The full Fourier space $\mathbb F_3^6$ contains
$(3^6-1)/(3-1)=364$ nonzero one-dimensional frequency lines.
Exactly $\binom62=15$ are permitted pair-difference lines,
leaving $349$ forbidden. Of those forbidden lines, $243$
are not invariant under $x\mapsto x+c\mathbf1$; they disappear
once the global-shift identifications are imposed at the
preparation level. The remaining quotient dual space has
$(3^5-1)/2=121$ nonzero lines, of which $106$ are forbidden.
For each forbidden quotient line $[r]$, the three uniform mixtures
conditioned on $r\cdot x=k$ contain $81$ distinct phase rays each,
and obey ideally
\begin{equation}
\overline\rho_{r,k}
=\frac1{81}\sum_{\substack{[x]\in\mathbb F_3^6/
\langle\mathbf1\rangle\\r\cdot x=k}}
\rho_x
=\frac{I_6}{6},\qquad k\in\mathbb F_3.
\label{eq:s9-mixtures}
\end{equation}
Here $r\cdot\mathbf1=0$, so the affine-class condition is well
defined on the quotient. Thus three matching settings do not mean
that only three preparation equivalences must be checked.

A subtle issue is measurement completeness. For the ideal
three-matching POVMs, the nine edges of $K_{3,3}$ span
$18$ independent off-diagonal Hermitian directions. Their
endpoint sums span five diagonal directions, including the
identity, so these measurements yield exactly $22$ independent
probabilities on normalized six-mode states, whereas a general
six-dimensional density operator has $6^2-1=35$ real parameters.
Consequently, these three witness settings alone cannot establish
full six-mode tomographic equivalence. Auxiliary fiducial measurements,
or well-justified assumptions restricting the accessible preparation
model, are needed to establish the mixture equivalences in the
intended operational theory. With experimental imperfections,
convexly mixed secondary preparations may be used to implement
exact equivalences, as demonstrated in other contextuality
experiments~\cite{mazurek}. Such a construction must be verified
for the present family of $106$ quotient constraints; its existence
should not be presumed. Alternative robust preparation-noncontextual
tests can avoid exact specified equivalences in other operational
scenarios~\cite{Pusey2018}, but do not directly supply a corrected
bound for this particular witness.

The ideal three-setting bound $P_{\NC}^{(3)}=8/9$ and visibility
criterion $v>2/3$ apply when the stipulated equivalences hold
exactly, as they do for the depolarized theoretical phase ensemble.
Measured approximate agreement of mixture statistics, by itself,
does not justify imposing those exact ontic constraints. A practical
test should either construct statistically certified secondary
preparations satisfying the requisite equivalences, or derive an
appropriate robustness statement for explicitly quantified
operational deviations. Confidence intervals for the exclusion
score, source calibration, relative phases, mode-dependent loss,
matching selection, and outcome-detection efficiencies must be
propagated consistently. The edge-resolved measurement witness is
separate: it requires a characterized source and does not follow
from a preparation-noncontextuality violation alone. Recent
experimental POVM-certification methods illustrate ways to account
for state-preparation imperfections in that distinct resource
test~\cite{Brinster2026}.

\end{document}